\documentclass[10pt,onecolumn,twoside]{IEEEtran} %!PN

\usepackage{multirow}
\usepackage{amsfonts}
\usepackage{epsfig}
\usepackage{amsmath}
\usepackage{amssymb}
\usepackage[nolist]{acronym}
\usepackage[english]{babel}
\usepackage{cite}
\usepackage{color}
\usepackage{stfloats}
\usepackage{algorithm}
\usepackage{algorithmic}
\usepackage{subfigure}

\newcommand{\rhot} {\rho_{\text{T}}}
\newcommand{\rhos} {\rho_{\text{S}}}

\newcommand{\Tdts}{ T_{\text{TS},l} }
\newcommand{\TA}{ T_{\text{A},l} }

\newcommand{\mat}{\pmb}
\newcommand{\TD}{T_{\text{D}}}

\begin{document}

\begin{acronym}
\acro{QoS}{quality-of-service } \acro{MSE}{mean square error }
\acro{MDC}{Multi-description codes} \acro{MRC}{multi-resolution
codes} \acro{RV}{random variable} \acro{GOP}{group of picture}
\acro{RCPC}{rate-compatible punctured convolutional}
\acro{SLV}[PLV]{packet loss visibility} \acro{OFDM}{orthogonal
frequency division multiplexing} \acro{CRC}{cyclic redundancy check}
\acro{EEP}{equal error protection} \acro{UEP}{unequal error
protection} \acro{FEC}{forward error correction} \acro{RB}{resource
block} \acro{CSI}{channel state information} \acro{SNR}
{signal-to-noise ratio} \acro{MCEC}{Motion-Compensated Error
Concealment} \acro{FR}{Full-reference} \acro{VQM}{Video Quality
Metric} \acro{PSAM}{pilot symbol assisted modulation}
\acro{NC}{network coding} \acro{JSCC}{joint source channel coding}
\acro{AWGN}{additive white gaussian noise} \acro{UCSD}{University of
California at San Diego} \acro{UGC}{user-generated content}
\acro{PDA}{Personal Digital Assistant} \acro{TDMA}{Time Division
Multiple Access} \acro{QeA}{question-and-answer} \acro{RD}{rate
distortion} \acro{QoE} {quality of experience} \acro{NC} {network
coding} \acro{MDP}{Markov decision process} \acro{MOS}{mean opinion
score} \acro{PSNR}{peak signal to noise ratio} \acro{RNC}{random
network coding} \acro{RS}{Reed-Solomon} \acro{MD}{multiple
description} \acro{SW}{Slepian-Wolf} \acro{WZ}{Wyner-Ziv}
\acro{SI}{side information} \acro{DSC}{distributed source coding}
\acro{SLEP}{Systematic Lossy Error Protection} \acro{PLR}{packet
loss rate} \acro{PRISM}{Power-efficient, Robust, high  compression,
Syndrome-based Multimedia} \acro{HARQ}{hybrid automatic repeat
request} \acro{MAC}{multiple access control}
\acro{MIMO}{multiple-input multiple-output} \acro{FEP}{forward error
protection} \acro{DP}{dynamic programming} \acro{MV}{multiview}
\acro{RADIO}{RAte DIstortion Optimization} \acro{WMSN}{wireless
multimedia sensor network} \acro{AP}{access point} \acro{DU}{data
unit}\acro{RA}{rate allocation}
\acro{FDMA}{frequency-division multiple access} 
\acro{ISA}{iterative sensitivity adjustment}
\acro{DVC} {distributed video coding}
\acro{MAC} {Medium Access Control}
\acro{PCF}{Point Coordination Function} 
\end{acronym}

\title{Multi-View Video Packet Scheduling}

\author{ Laura  Toni,~\IEEEmembership{Member,~IEEE},  Thomas Maugey,~\IEEEmembership{Member,~IEEE},   and Pascal Frossard,~\IEEEmembership{Senior Member,~IEEE} \thanks{L. Toni, T. Maugey and P. Frossard are with \'Ecole Polytechnique F\'ed\'erale de Lausanne (EPFL), Signal Processing Laboratory - LTS4, CH-1015 Lausanne, Switzerland. Email: \{laura.toni, thomas.maugey, pascal.frossard\}@epfl.ch.} \\
}
\maketitle
\thispagestyle{empty}
%%%%%%%%%%%%%%%%%%%%%%%%%%%%%%%%%%%%%%%%%%%%%%%%%%%%%%%%%%%%%%%%%%%%%%
\begin{abstract}
In multiview applications, multiple cameras acquire the same scene from different
viewpoints and generally produce correlated video streams. This results in large amounts of highly redundant data. In order to save resources, it is critical to handle properly  this correlation during encoding and transmission of the multiview data.
 In this work, we propose a correlation-aware packet scheduling algorithm for multi-camera networks, where information from all cameras are transmitted over a bottleneck channel to clients that reconstruct  the  multiview images. The scheduling algorithm relies on a new rate-distortion model that captures the importance of each view in the scene reconstruction. We   propose a problem formulation for the  optimization of the packet scheduling policies, which  adapt to   variations in the scene content. Then, we design  a low complexity    scheduling algorithm based on  a trellis search that selects the subset of candidate packets to be transmitted towards effective multiview reconstruction at clients. Extensive simulation results confirm  the gain of our  scheduling algorithm when  inter-source  correlation information is used  in   the scheduler, compared to scheduling policies with no information about the correlation or non-adaptive scheduling policies.   We finally show that increasing the optimization horizon in the packet scheduling algorithm improves the transmission performance, especially in   scenarios 
 where the level of correlation rapidly varies   with time.
 \end{abstract}
%%%%%%%%%%%%%%%%%%%%%%%%%%%%%%%%%%%%%%%%%%%%%%%%%%%%%%%%%%%%%%%%%%%%%%
\section{Introduction}
Advances in   interactive services and   3D television have paved the road to multiview video applications, in
which multiple sources acquire and transmit several correlated  media
streams~\cite{YiuJinChan:J07,CheOrtChe:J11,MauFro:J12,KubSmoMagCheZha:J07}.
Multimedia
wireless sensor networks and multi-camera video systems
%acquisitions, in
%which several cameras acquire the same scene from different
%perspectives, 
are typical examples of multiview setups. The flexibility and the interactivity offered by such
applications however come   at the price of   increased storage/bandwidth
requirements.  To overcome these limitations, the coding and transmission schemes need to 
properly exploit the correlation among sources, in order to provide effective image quality in resources constrained environments.

 In this context, we aim at providing insights on how resource allocation strategies can benefit from correlation information in a multi-camera  scenario, in which    neighboring cameras acquire the same scene but from different perspectives. This   scenario results in  spatial correlation between the information streams, since cameras typically have overlapping fields of view, in addition to temporal correlation between frames acquired consecutively by the   same camera. This spatial-temporal correlation can be exploited either at the source (e.g., by joint encoding of the different sources) or at the decoder side (e.g., by joint reconstruction of the different images).  In this work, we consider the latter case and we  show  how the packet transmission scheme can be opportunistically adapted to satisfy network constraints, when the source correlation is  exploited at the decoder for image   reconstruction.

  In more details, the proposed framework targets the optimization of resource allocation schemes for the transmission of  correlated sources under delay and bandwidth constraints. Rather than focusing here on   source coding  aspects, we are interested in a   scenario where each camera  independently acquires part of a scene with no communication  between cameras. The encoded views need to be gathered by a gateway or a wireless \ac{AP} (see Fig. 1), which then forwards packets to clients  interested in decoding (part of) the 3D scene. Assuming that network resources are constrained, only a subset of the camera images can be transmitted to the clients.    The encoded views   are   transmitted with a \emph{correlation-aware packet scheduling} algorithm driven by the gateway or the \ac{AP}.
 This centrally coordinated  scenario is quite typical in practice, and in particular   in IEEE 802.11 wireless networks. In these networks, the \ac{PCF} is one of the common solutions supported by \ac{MAC} layer to organize data transmission \cite{IEEESTD:12,Jia_Pat:04}.
 { At higher layers, master routers or home gateway devices are also  used as central  controllers for  network services and  devices \cite{Cisco:WP09,TI:WP05}.  }
% , in which decision strategies are usually performed at a central node. The \ac{MAC} layer, for example, supports as   solution  to schedule the optimal access mechanism   the \ac{PCF}. The latter   is an access method  based on polling scheme controlled by a   point coordinator, which is usually the AP  \cite{IEEESTD:12,Jia_Pat:04}. Looking at higher layers, where we actually consider our packet scheduling strategy, a centralized structure for packet routing, packet filtering, etc.  is also considered \cite{Capozzi}. Master routers or home gateway devices are   used as controller for either network services and network devices \cite{Cisco:WP09,TI:WP05}.  
%%%%%%%%%%%%%%
\begin{figure}[t]
\begin{center}
\includegraphics[width=0.9\textwidth,  draft=false]{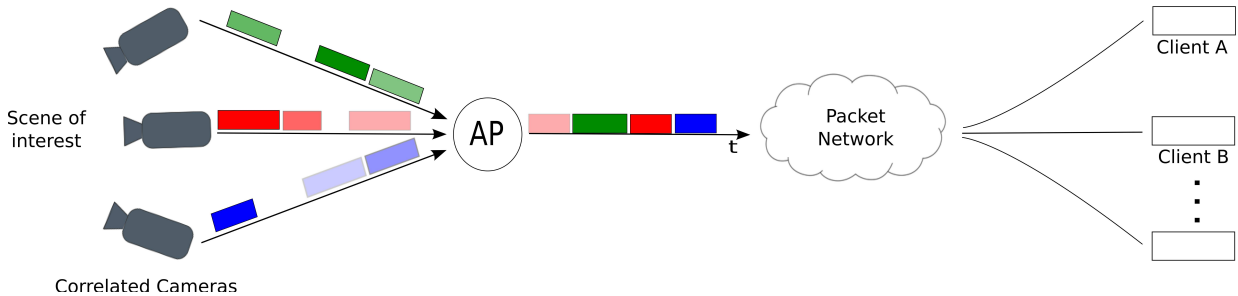}
\caption{Multi-camera system, with bandwidth bottleneck at the access point.}\label{fig:Scenario_PF}
\end{center}
\end{figure}
%%%%%%%%%%%%%%%%%%%%%
The packet scheduling algorithm filters packets to reduce the transmission cost and satisfy the resource constraints in the system under the assumption that the images are jointly reconstructed at decoder.   
  In order to optimize the reconstruction quality, one has however to properly select the    packets to be transmitted, along with their transmission schedule.
 For example, the frames that are highly correlated to packets already available at the decoder can have a low priority in the scheduling algorithm. This is due to the fact that they can be   reconstructed  from the correlated frames at the decoder side even if they are actually not transmitted. On the other hand,   frames that have   only a low correlation with previously transmitted data  should be prioritized in the scheduling since they would be reconstructed at a poor quality if they are not transmitted. 
 
We propose a novel \ac{RD} model that estimates the distortion in scene reconstruction from multiple correlated images.
Based on this model,  we build a   scheduling technique that  minimizes the distortion in the scene reconstruction  and   adapts   the transmission scheme to temporal variations of the scene content and   correlation level.     
The  proposed scheduling algorithm optimizes the long-term utility function with refinement at  each transmission opportunity. For such an algorithm to reach optimality though, a large time horizon has to be considered in the optimization, which leads to high computational complexity. Thus, we propose a suboptimal trellis-based   algorithm that is  able to reduce the complexity while still preserving most of the benefits of correlation-aware scheduling optimization.   
 Simulation results demonstrate that the proposed scheduling algorithm  outperforms  correlation-agnostic  scheduling policies  or static camera selection algorithms.
  This shows    the need of   correlation-aware scheduling  policies in multiviews systems, which are able to efficiently share network resources among cameras, while   \ac{RA} techniques proposed in the literature cannot   solve such a scheduling  problem, since they usually do not consider    correlation between sources~\cite{ZhoChe:J11,ShiSch:J10}.

The remainder of this paper is organized as follows.   Related works on multiview data gathering 
are described in Section  \ref{sec:related_works}.  In Section
\ref{sec:frameworks},   some technical preliminaries are given   and our new \ac{RD} model is introduced. The packet scheduling problem is
formulated in Section \ref{sec:packet_scheduling} {and the trellis-based optimization solution is provided in Section \ref{sec:trellis}}.   In
Section \ref{sec:results}, we discuss the  simulation results, and we conclude in Section  \ref{sec:conclusions}. 

%%%%%%%%%%%%%%%%%%%%%%%%%%%%%%%%%%%%%%%%%%%%%%%%
\section{ {Related Works}}
\label{sec:related_works}   In this section, we first provide a general overview of the most
relevant works from the literature that focus  on multi-camera
streaming and we highlight the key differences with our work.  Then, we describe in more detail the research work in  resource allocation and correlation-aware multiview streaming.

In  multiview systems, prior studies usually addressed two main open problems: i) how to efficiently encode distributed sources, ii) how to efficiently deliver information to users in different applications. To answer the first question,  \ac{DSC} 
%which is based on the \ac{SW} \cite{SleWolf:J73} and \ac{WZ} \cite{WynZiv:J76} information theory theorems,  
  has gained attention as new coding paradigm \cite{SleWolf:J73,WynZiv:J76} to exploit source correlation. When no communication is assumed between cameras during the coding process, \ac{DSC}  allows the encoding to stay  simple by shifting the computational complexity   to the decoder. Research on DSC, as well on   \ac{DVC}, has been mainly focused on optimizing the coding scheme, given an a priori knowledge on the correlation, i.e., given  an a priori    \ac{SI}~\cite{XioLivChe:J04,WuStaXioKun:J09,springerlink:11}.
Thus,   the  selection   of     sources that can be used for the generation of  {SI} is usually assumed to be known; the optimization of  this selection  is still an open problem.  Even if many works have studied DSC in multiview applications,   an optimization framework that is able to   exploit in the most efficient way the source  correlation level is still missing. In our paper, similarly to the DSC framework, we consider that the cameras do not communicate with each other but rather exploit the source correlation in the packet scheduling process. Even if this is not considered in this paper, our framework also applies to cameras streams encoded by DSC. It represents a complementary solution to DSC in the design of distributed camera systems.

% It also requires high computational complexity at encoder and possibly expensive communication between sources. 
%In this context,%Several efforts have been made to design   practical \ac{DSC}-based
%communication schemes for correlated information sources~\cite{XioLivChe:J04,WuStaXioKun:J09,springerlink:11}.
% However, even if \ac{DSC} permits to reduce bandwidth requirements,
 %high-complexity decoding schemes are generally induced,   encoders  usually  require some a priori information about the correlation between sources, and  the application of DSC to many sources  rapidly reaches complexity limits. A  \ac{DVC} has been proposed in~\cite{AnaAgrBull:C08} for multiview sequences, where the correlation among adjacent cameras is exploited  to construct the \ac{SI} more accurately, while    a clustered coding strategy has been proposed in \cite{WanDaiAky:J11} for wireless multimedia sensor networks.
%Even if the  coding   complexity is reduced in this case, 
%\ac{DSC}   still depends heavily on a good knowledge of correlation at sources. In multiview scenarios, the   shape of this correlation has not been  properly modeled yet. Thus, although DSC relies on a priori selection on which sources can be used for the generation of \ac{SI}, the optimization of this selection policy is still an open problem.  
%\new{Thus, in a distributed source  scenario, it is still unclear how to exploit the source correlation in the coding-transmission-decoding chain.}

In the second set of works that optimize the delivery of multiview data, some   prior  studies    address  the problem of providing interactivity in selecting views, while saving on transmitted bandwidth and view-switching delay~\cite{KurCivTek:J07, CheVelOrt:J11, LiuQinMaZha:J10, CheOrtChe:J11,KueSik:C08,CheZhaSunShi:C09}. The work in~\cite{CheOrtChe:J11} is mainly focused on coding views with a minimum level of redundancy in order to simplify the view switching, and the works in~\cite{LiuQinMaZha:J10,PanIkuBanWat:C11}   optimize  the selection of views to be encoded and transmitted based on the user interest. The authors in~\cite{KueSik:C08,LouGuaHuaJia:J07}   investigate  the transmission of  multiview video coded streams on P2P networks and IP multicast, respectively. These works mainly focus on the coding aspects and \ac{DSC} is often proposed as a solution to reduce encoding complexity \cite{AnaAgrBull:C08} or to provide interactive access to the different views \cite{CheOrtChe:C09}.

The work proposed in this paper is rather defined as a rate allocation problem in multi-camera systems. Multi-camera resource allocation solutions in the literature often ignore the dynamic correlation between sources  and rather focus on optimizing the resources for  each camera independently. In other words, they  usually optimize the
scheduling policy in evaluating the cost, the distortion gain and the
time constraints of each camera separately and ignores the possible correlation
among cameras.  This may result in suboptimal allocation of the
network resources.       Resource allocation
techniques have for example been considered  in~\cite{ShiSch:J10} for video surveillance systems, in
which each of the camera captures and transmits the
video information in a multihop network. The optimization of the 
resource allocation (i.e., the time sharing between sources) is
based on both the network and source information, but ignores  the correlation
between the sources.  In a more general   resource allocation framework, few works have introduced the sources correlation in the  optimization of transmission schemes.  {A multi-party 3D tele-immersive system is considered in \cite{Yang:2010}, where correlated views are rendered together to create a common virtual environment among all   participants. These participants are distributed over an overlay network and   can gather information from neighboring nodes. Source correlation is taken into account to dynamically optimize the multicast topology for content delivery between nodes involved into   the multi-party 3D tele-immersive session. }
In~\cite{Dorna:J11},  a three-step approach is proposed to optimize the resource allocation between spatially correlated sources  for multi-cell \ac{FDMA} networks. However, multimedia transmission is not considered in the optimization.

In~\cite{VurAky:J06}, the level of spatial correlation between sources has been
considered at the \ac{MAC} layer for   wireless sensor networks. The authors assume that the network needs to estimate an
event $S$. Due to the correlation between neighboring sensors,
only part of them might be selected for sending information to the
sink, so that the transmission data rate is limited. The MAC
protocol  prioritizes the  access to representative nodes, i.e.,
nodes with reduced levels of correlation. The same intuition has
been considered in~\cite{DaiAky:J09} and applied to multimedia
streaming.  A spatial correlation model for visual information in \acp{WMSN}  has been proposed, introducing an entropy-based analytical framework to evaluate the visual information offered by
 multiple cameras. When the network resources are insufficient the cameras that  maximize the joint entropy in a  camera set are selected for transmission. 
 The model however only solves a static
correlation-based \textit{camera selection} technique, while  we consider a dynamic correlation-based \textit{packet scheduling}
optimization in our work. In particular, the framework in~\cite{DaiAky:J09} can be seen as a particular case of our problem, where both cameras and scene content are static.
  The correlation model proposed in~\cite{DaiAky:J09} has been also  used in~\cite{WanDaiAky:C11}, where the problem of efficient gathering of visually correlated images from multiple sensors has been investigated.  %The contribution of the author is threefolds: i) the optimal location to place multimedia processing hubs is found in order to ensure the  effectiveness of channels frequency reuse; ii) the grouping of  cameras   into    hubs has been optimized  in such a way that the joint compression gain is maximized by jointly encoding correlated images associated  to the same hub; iii) a scheduling optimization is proposed with the final  goal of network lifetime maximization.
%Each sensor differentially  encodes its image conditionally to previously overheard transmissions of broadcaster nodes.  
The scheduling optimization is aimed at reducing the energy consumption during    transmissions by exploiting a correlation-aware differential encoding technique. However, the model is highly sensitive to transmission failures. Moreover, the cameras grouping optimization is based on the assumption of a static correlation, which does not hold in dynamic scenarios. Our work is substantially different from~\cite{WanDaiAky:C11}, since we propose a packet scheduling optimization that  i)  is able to   adapt to correlation variations in dynamic scenes,  ii) considers  independent source  coding (i.e., it preserves  simplicity at the source side). 
 
Finally, it is worth noting that the
correlation between cameras might be exploited not only for DSC or resource allocation techniques, but also for error resilience. For example, the correlation between views is implicitly considered in~\cite{LiuCheJi:C11}. The authors   propose an optimized interactive multiview streaming over wireless wide area networks  (WWAN), where a cooperative peer-to-peer repair  technique is considered to alleviate packet losses.%; the transmission decision of each peer during CPR is optimized based on a \ac{MDP}   that exploits the view correlation in data recovery. Even if view recovery through correlated  neighbors is considered in~\cite{LiuCheJi:C11}, the essence of this study is radically different from our work, in which we have no multicast session, no interactive streaming but only multiple cameras that share   limited network resources  and transmit correlated video streams to a central node. 
%Multiview data protection from channel
%impairments has also been investigated in \cite{LiuCheVelEkrYus:C10},
%where a cooperative peer-to-peer repair strategy is proposed
%for clients watching different views. Assuming that not only a texture
%map but also a depth map is transmitted, each node is able to
%repair the packet losses of another node, that receives another camera view. This method exploits the correlation between multiple cameras 
%as a way to combat channel impairments. 
%%It addresses the joint source/channel coding optimization and evaluates the best redundancy level that should be considered for the depth
%%map and the FEC channel coding.
%Another  error resilience method based on  redundant coding has been
%proposed in \cite{DisSilWorFer:C10} for \ac{MV} systems, in which
%both   temporal and view correlations are exploited for error resiliency. 
%%

There are important differences between the above works and the study proposed in this paper. First, we focus our attention on the important problem of optimizing
  scheduling algorithms such that view correlation can be exploited efficiently at the decoder.
 Second, even if some other works have investigated resource allocation
techniques for multiview scenarios, dynamic view correlation and dynamic packet scheduling solutions are not studied   in the literature related to multi-camera systems. This is exactly what we propose to address in this paper.  
%
%is not  considered in the problem formulation,  and camera
%selection algorithms have been studied rather than dynamic packet scheduling
%techniques algorithms as considered in this work.  

\section{Framework}
\label{sec:frameworks}
We now describe the framework considered in our work. First, we present  the multi-camera system and describe the multiview acquisition and transmission processes. Then, we introduce  the scene reconstruction method and show that the correlation between cameras plays a crucial role in the reconstruction of missing frames  at the decoder. Finally, we propose a new rate-distortion model for the representation of the 3D scene information.

\subsection{Multi-camera system}
We consider $M$ cameras that acquire images and depth information of a 3D scene from different viewpoints. The images acquired by the $M$ correlated cameras need to be collected by  a common \ac{AP} that eventually transmits (part of) the   3D scene information to clients, which are all interested in receiving all video streams. 
Due to bandwidth constraints   in the communication system (e.g., on the wireless channel, or on the path between AP and clients), it might not be possible to transmit all the frames from all the cameras to the clients.  Thus, at each transmission opportunity, it is important to accurately select which images have to be scheduled and which ones can be sacrificed (i.e., not transmitted), such that the average distortion is minimized. However, depending on the camera arrangement and the scene information, the frames acquired from the different cameras might be correlated
in both time and space. First, each camera acquires temporally consecutive
frames, which are correlated, especially for static or low-motion 3D scenes: this is the \textit{temporal correlation}   in image sequences.
Then, neighbouring cameras might acquire  overlapping
portions of the same scene; this   leads to correlated  frames due to  the \textit{spatial correlation} between multiview cameras. Both the temporal and the spatial correlations  might help in reconstructing the overall scene information  if some images are missing at the decoder.
 %Since this level of correlation might help in reconstructing the scene information even if some views are missing,
%it might be important to exploit this correlation level in our communication problem.
%In particular, we consider the case in which the

 We address the frame selection problem as a resource allocation problem that takes into account the level of correlation among cameras in a novel packet scheduling algorithm.  We assume a model in which there is no communication among cameras  in order to save bandwidth and power. The only minimal  information that is known a priori is the position of the cameras, which is possibly updated when cameras change positions in dynamic settings.  Along with  depth information, each camera is   able to estimate its influence on its neighbors and in particular the contribution that it can offer in the reconstruction of neighbor views. We propose below a novel correlation model where each camera can predict the correlation level with neighboring cameras, without global depth information. This   local  correlation level, which is a set of simple values representing the influence of the camera in the reconstruction of the neighboring ones,  is sent by each camera to the scheduling engine.

Then, we consider that each encoded  image at
a given time  instant from a particular  camera is packetized into a \ac{DU} and
stored in the camera buffer. Each data unit contains texture and depth information about the 3D scene. All the  camera  \acp{DU} are possible candidates for scheduling.
 We further assume that the transmission  is based on a \ac{TDMA} model where no more than one DU might be scheduled  in any \ac{TDMA} slot. Once
a DU is scheduled for transmission, the channel stays busy for one or multiple time slots, until   the
current DU has been completely transmitted.\footnote{From here onwards, we assume  the  time axis discretized  in slots (or scheduling
slots)  of length equal to the TDMA slot duration.}
 Due to  streaming delay
constraints,  the \ac{DU}  needs to be received before a playback  deadline, denoted by $\TD$, in order to be useful for decoding.
 This means that a   DU acquired at the time
$t$  stays useful till   time $t+\TD$. Data units that have no chance to be received on time are not considered for   scheduling and simply dropped by the cameras. 
 We also assume  that the communication channel is lossless such that
 all the transmitted \acp{DU} are   correctly received by the access point and subsequently the clients.  It follows that packets that are not available at decoder have been skipped by the scheduler, and not lost due to unreliable communication.
  In this framework, our goal is to propose a correlation-aware scheduling algorithm that selects DUs from different cameras in such a way that the overall distortion in the reconstruction of all camera views is minimized under the   bandwidth constraints. 
  
\subsection{Scene Reconstruction} 
\label{subsec:RD}
We describe now  the scene reconstruction process, which will help to better understand the benefits of exploiting the spatial and temporal correlation of the images. 
At the receiver side, each 
frame is decoded independently. The images that have not been transmitted are
estimated based on time and/or view interpolation algorithms using information from neighboring frames.
\begin{figure}
\begin{center}
 {\subfigure[ ]{
\includegraphics[width=0.65\linewidth,  draft=false]{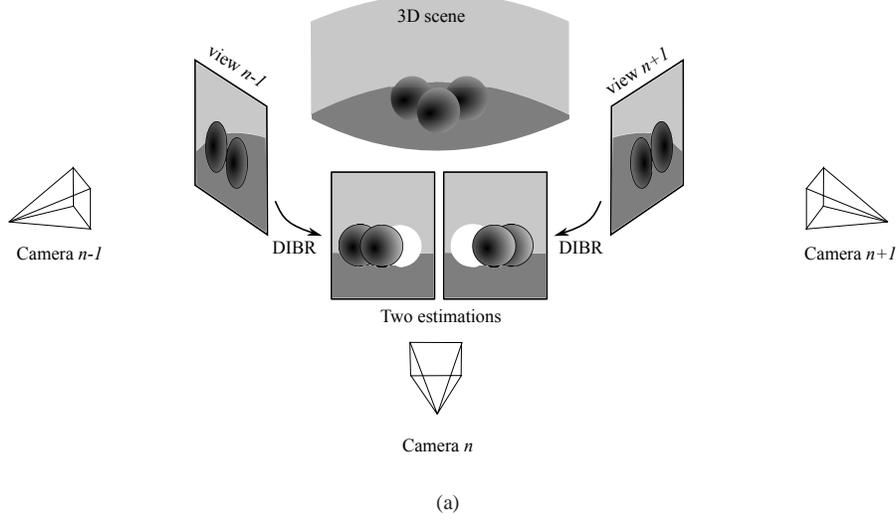}
\label{fig:DIBR_A} }}  \hfill   
\end{center}
\begin{center}
{ \subfigure[]{
\includegraphics[width=0.55\linewidth,  draft=false]{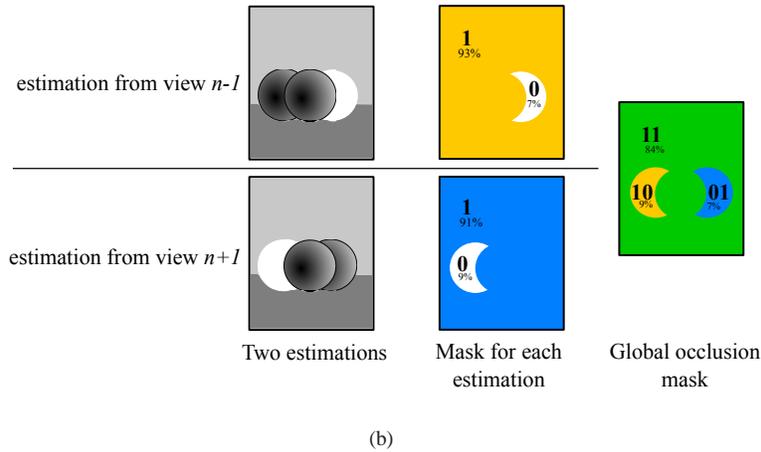}
\label{fig:DIBR_B} }}   \caption{Example of DIBR image estimation at decoder. (a)
the central  view $n$ is estimated from the two neighboring
views $n-1$  and $n+1$. (b) the occlusion maps corresponding to
the two  estimations are merged in order to obtain a global occlusion map with 3 regions.  The percentage numbers  in the masks indicate  the  portion of the frame dedicated to each region.} \label{fig:DIBR}
\end{center}
\end{figure}
 More precisely, for the interpolation of a missing  view $n$, the receiver uses images from   neighboring cameras with help of
  depth image based rendering (DIBR) techniques (Fig.~\ref{fig:DIBR_A}). Typically, DIBR algorithms use  depth 
information in order to  estimate by projection the position of pixels from view $k$ in  the missing view $n$. The   projected pixels
are generally of  good precision (depending on the accuracy of the depth map
\cite{Muller_K_2011_pieee_tdv_rudm}) but do not cover the whole
estimated image, due to visual occlusions. As
shown in Fig.~\ref{fig:DIBR_B}, one can build a binary mask that describes the occluded regions. Then, by  merging the estimations obtained by the projections of
different neighboring  views, we obtain different reconstructed regions in the interpolated image. 
This can be summarized in a global occlusion map with different regions corresponding to the different occlusions.  
 In the example in Fig.~\ref{fig:DIBR_B}, the reconstructed scene is subdivided into three regions, each of them is characterized by the set of neighboring views that contribute to the scene reconstruction. In particular, the blue region (which represents   $7\%$ of the total scene) is reconstructed based on the  estimation from only the view $n+1$, while for the yellow one (which represents   $9\%$ of the total scene) the estimation from view $n-1$ is considered. The remaining $84\%$ of the scene (i.e., the green region) is reconstructed by merging estimations from both views. 
The principle for temporal extrapolation is the same. The decoder uses the available past
frames to reconstruct a missing frame. The past frames cannot be used to estimate the
whole missing image because of occlusions and object motion. The regions where the past frames could
give some  useful information are computed similarly to the occlusion map in the view  interpolation case. The global map with the different prediction regions is used to decide on the best interpolation   method for the   missing frames at the decoder.
  
\begin{figure}[t]
\begin{center}
\includegraphics[width=0.60\linewidth,draft=false]{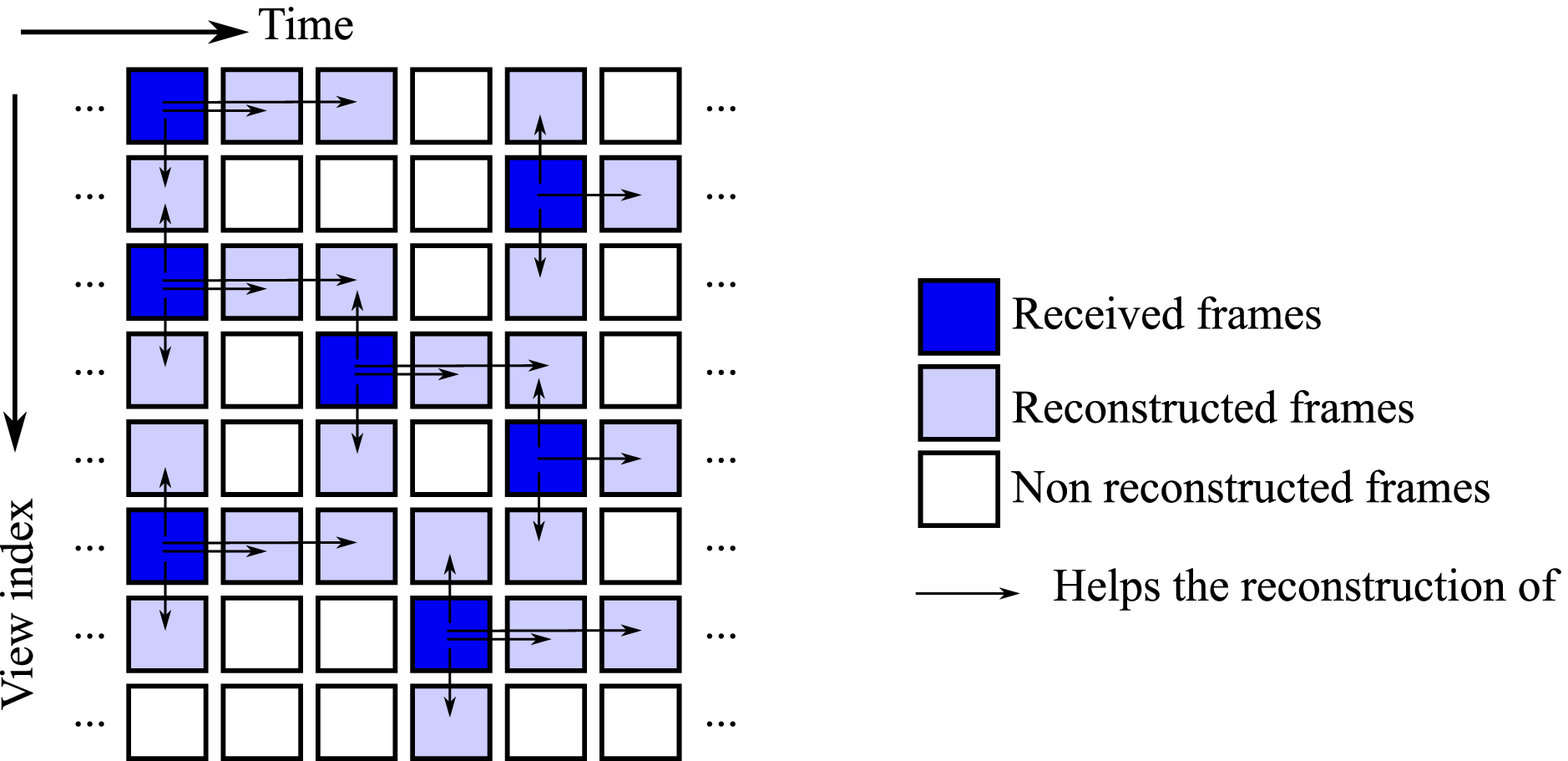}
 \caption{Example of frames reconstruction in multiview video setup, where each frame is correlated with the frames of  two neighboring  views and with the two temporally consecutive frames  (of the same view). Missing frames are reconstructed from information in the correlated frames that are available at decoder.   Received frames are represented in the figure by dark colored boxes, the reconstructed ones by light colored boxes. White boxes represent frames that cannot be reconstructed from the received frames.}\label{fig:FrameReconstruction}
\end{center}
\end{figure}

An example of multiview video
reconstruction is depicted in Fig. \ref{fig:FrameReconstruction},
for the case of $8$ cameras that acquire  several temporally consecutive frames.
The goal of the decoder is to reconstruct all the frames in
time and space, even if only part of them have been received
(dark colored boxes in Fig. \ref{fig:FrameReconstruction}). In this example, we consider that  each frame  is correlated with frames of the  two neighboring views in space, and with the  two temporally successive frames (of the same view). If one or more of these correlated frames
are missing, the received frames can contribute to the estimation
of the missing data (light colored boxes in Fig. \ref{fig:FrameReconstruction}). In order to avoid error propagation, we consider that only the received frames can be used to reconstruct the missing ones (i.e., reconstructed frames are never used for   estimation of other missing frames). Note that we consider temporal estimation only in the forward direction for the sake of simplicity. Our model can however be extended easily to include temporal interpolation in the backward direction too (i.e., from future frames). Finally, a missing frame  
cannot be reconstructed (white boxes in Fig. \ref{fig:FrameReconstruction}) when all its correlated frames are missing too.

\subsection{Rate-Distortion Model}
\label{sec:RD}
We  now   propose a novel rate-distortion model adapted to the scene reconstruction framework described above.    The $m$-th camera at time  $t$,   acquires the image $F_{t,m}$ and compresses it at a rate of $R_{t,m}$ bits per pixel $
(m=1,\ldots, M)$.  A subset of the compressed images captured by all cameras is 
   transmitted to the decoder, which targets the reconstruction of the full scene.  
If  the frame $F_{t,m}$ is  available at the decoder,
the   distortion is directly dependent on the compression or the source rate.
If $F_{t,m}$ is missing at decoder, it is   reconstructed from the available neighboring frames (in time and space), as described in the previous section.

The    overall distortion of the scene   at  
instant $t$   is thus expressed as
\begin{equation}\label{eq:RD_1}
D_{t}(\mat{R}_t)=\sum_{m=1}^{M} \frac{1}{w_m} \ D_{t,m} (\mat{R}_{t})
\end{equation}
where  $w_m$ represents the  relative importance of a given   camera view. It permits to give a different weight to   each  camera view in the   distortion evaluation (e.g.,  the central camera might be preferred to the lateral
ones)  and it reflects the relative interest that clients have in each camera stream. In our problem formulation, the weight parameter is assumed to be   given as a priori information. 
The rate vector $\mat{R}_{t}$, defined as
$$\mat{R}_{t}=[R_{t,1} \ R_{t,2} \ldots
R_{t,M}  \  R_{t-1,1} \ldots R_{t-1,M} \ \ldots R_{t-\rhot,1}
\ldots R_{t-\rhot,M}]^T\,,$$
  represents the size (in bpp) of
the frames received from the different cameras $(m=1,\ldots, M)$ in a window of time of size $\rhot$, which can be used for the reconstruction of $F_{t,m}$. The parameter $\rhot$ defines the maximum number of frames that can be considered in temporal interpolation at the decoder. %Thus, if  the frame $F_{t,m}$ is missing at the decoder side, it can be reconstructed  from the neighboring views in space or by the $\rhot$ previously acquired. 
 The distortion
  $D_{t,m} (\mat{R}_{t})$ is the distortion of the $m$-th view at instant $t$. For each view  $m$ acquired at
the instant $t$, we further decompose the frame into regions $s_j$   and we denote by ${\mathcal{S}}_{t,m}$ the set of such regions. 
 For each $s_j \in {\mathcal{S}}_{t,m}$, we denote by $\alpha(s_j)$  the relative area of the frame dedicated to
the region $s_j$, such that $\sum_{s_j \in {\mathcal{S}}_{t,m}} \alpha(s_j)=1$.  
In Fig.~\ref{fig:DIBR}, for example, the frame acquired from the  central camera  is  subdivided in three different regions: the blue, the yellow, and the green ones, with $ \alpha(s_j)$ corresponding to $0.07, 0.09,$ and $0.84$ respectively.

 Then, a mapping function $\mat{\phi}_{j,m,t}$ describes which of the neighboring frames  can contribute  to the reconstruction of the region $s_j$ of the $m$-th view at time $t$. In  the absence of temporal correlation,  the
spatially neighboring views only are considered for frame reconstruction.
This means that $\mat{\phi}_{j,m,t}=[\phi_{j,m,t}(1) \ldots
\phi_{j,m,t}(M)]$, where  $\phi_{j,m,t}(k)=1$ if the $k$-th camera
is correlated with the region $s_j$ of the frame $F_{t,m}$ and $\phi_{j,m,t}(k)=0$ otherwise. In this case, $\mat{R}_{t}$ reduces to $\mat{R}_{t}=[R_{t,1} \ R_{t,2} \ldots
R_{t,M}]$.  When both spatial and temporal correlations are used in the reconstruction,  the matrix $\mat{\phi}_{j,m,t}$  becomes
%\begin{align}
$$
\mat{\phi}_{j,m,t}  =  [\phi_{j,m,t}(1) \ldots
\phi_{j,m,t}(M) \ \phi_{j,m,t-1}(1) \ldots \ \phi_{j,m,t-1}(M) \
\ldots \phi_{j,m,t-\rhot}(1)  \ \ldots
\phi_{j,m,t-\rhot}(M)]
$$
%\end{align}
where $\rhot$ is the number of past frames that can be considered
for the reconstruction of the current image.
Equipped with the above notation, the distortion $D_{t,m} (\mat{R}_{t})$ becomes the sum of the distortion in each part $s_j^{t}$ of the frame at instant $t$:
\begin{equation}\label{eq:RD_2}
D_{t,m} (\mat{R}_{t}) =  \left\{ \begin{array}{ll}
\sum_{s_j \in  {\mathcal{S}}_{t,m}  } \alpha(s_j) d\left[\mat{\phi}_{j,m,t}  \cdot  \mat{R}_{t}  \right] & \text{if the  view   is not received}\\
d\left[  {R}_{t,m}    \right]  & \text{otherwise.} \\
\end{array}
\right. 
\end{equation}
Finally, the   distortion functions $d[R]$ in Eq.~\eqref{eq:RD_2}  can be
evaluated from the general  expression of the \ac{RD}
function of an intra-coded frame  with   high-rate
assumption~\cite{CovTho:B91}:
\begin{equation}\label{eq:RD_3}
d[R_I]=\mu_I \sigma_I^2 \, 2^{-2 R_I }
\end{equation}
 where $R_I$ is the number of     bits per pixels and is equal to the sum of the rates that contribute to the current region,  $\sigma^2_I$ is the spatial variance of the frame   and $\mu_I$ is a constant
depending on the source distribution.  It is worth noting that the model of Eq. \eqref{eq:RD_3} has been chosen because   it is quite simple and yet accurate.  However,  our packet scheduling framework is general and other source rate-distortion functions could be used in Eq.  \eqref{eq:RD_2}.

\section{Packet Scheduling Algorithm}
\label{sec:packet_scheduling}
We  discuss in this section  a novel packet scheduling framework for wireless  multiview camera system that uses the rate-distortion model proposed in the previous section. Then, we   propose a novel problem formulation for  rate-distortion optimal packet  scheduling.
 
\subsection{Transmission policy}
We consider a channel with successive time slots for packet transmission. Each time slot represents a transmission opportunity. The objective is to select which \ac{DU} should be transmitted at each available time slot, in order to maximize the quality at the decoder  under the playback delay constraint given by $\TD$.  A greedy hence myopic   strategy   can choose  the    scheduling policy by selecting to transmit at each time slot the frame that minimizes the overall distortion at decoder. However,   such a scheduling solution does not necessarily optimize  the overall distortion since it does not consider a long term optimization objective.  A less myopic scheduling leads the scheduler to  allocate more fairly all the views of the camera set with  a more global distortion objective.  Thus, in the following we optimize the packet scheduling strategy over a finite time horizon that is generally larger than one transmission time slot. 

 The delay  $\TD$  as well as any   temporal parameter introduced in the following  is expressed in terms of time slots for the sake of clarity.  We denote  by $t$ the time slot at which we optimize the scheduling policy for a time horizon of $K$ time slots. We consider an online optimization with no a priori information about the video sequence. However, we allow a latency of $K$ slots between the acquisition and the scheduling process, in such a way that, at time $t$, the   characteristics of frames acquired up to the time slot $(t+K-1)$ are available to the scheduler. In more details, at 
  the time instant $t$, all the  frames  from all the views  acquired  
in the interval $[t-\TD+1,t+K-1]$ are possible candidates for transmission except those that have been scheduled already.  They form a set  of   cardinality $L$.  
Let the $l$-th DU be characterized
by its size  $B_l$ in  bits\footnote{The size of a DU includes the size of both  texture and depth data.},    its  acquisition  time slot $\TA$ (i.e., the instant at which the frame is
acquired), its expiration deadline $\Tdts=\TA+\TD$, and its transmission policy
$ \pi_l:\{a_l(1) \ldots a_l(K)\}$ in the next $K$ time slots.
A transmission policy $\pi_l$ at time $t$ is    
a binary vector according to which the DU $l$ is allocated  for transmission  over the  time
horizon   $[t, t+K-1]$. Let $\mathcal{A}=\{0,1\}$ be the action space and $a_l(k)\in \mathcal{A}$ the scheduling action   taken for the DU $l$ at the   $k$-th slot of  the optimization.
In particular,  $a_l(k)=1$ means that the data unit
$l$ has to be sent   at time $(t+k-1)$. 
 As the channel is lossless, we assume that each \ac{DU} is scheduled at most once during its lifetime and  that  each transmitted DU is sent entirely.  
In order to avoid transmitted DUs whose deadline has expired, we impose that at the $k$-th slot (with $k=1,\ldots, K$) only DUs acquired in the time interval $[t-\TD+k+1,t+K-1]$ are candidates for being transmitted  at  time     $(t+k)$.    Finally, we denote by $\mat{\pi}=[ {\pi}_1 \ldots {\pi}_L]^T$ the   scheduling policy for   the $L$ candidate DUs at time $t$. Each policy  $\mat{\pi}$ leads to a particular   distortion on the client side.  In this work, we seek   the best policy $\mat{\pi}^{\star}$  that is able to minimize the expected distortion while satisfying the channel constraints.  

The  scheduling policy is refined at next transmission opportunity based on the newly acquired frames. This means that a scheduling policy can change over time. In particular, among the best set of DUs selected for transmission, the DU scheduled in the first time slot is sent, while the scheduling is not guaranteed for the other DUs. For example,  a DU planned for transmission by the scheduling policy computed at time $t$ might actually never be  transmitted if a future  frame with higher importance  takes its transmission slot.    In this way, the refinement of the scheduling policy compensates for the limited knowledge of the video sequence that is imposed by the online nature of the algorithm.  We formally define below the packet scheduling problem in our new framework.

\subsection{Problem Formulation}
We first consider the scheduling problem for a single DU. In this case,    the   transmission  rate   is denoted by 
 $$ 
   {\mathcal{R}}\left({\pi_l} \right)   =   {B_l}   \left[ \sum_{k=1}^K a_l(k) \right]    
 $$   
 where $\sum_{k=1}^K a_l(k)$ is equal to $1$ is the DU $l$ is   scheduled for transmission in the $k$-th slot, and equal to $0$ otherwise.   The  overall  distortion is evaluated as 
\begin{equation}\label{eq:Dist_1DU}
   {\mathcal{D}}\left({\pi_l},    \mathcal{H}\right) =\left\{ \begin{array}{ll}
   D_{l}\left(\Psi\left\{\mathcal{H} \right\}   \right)  & \text{ if } \sum_{k=1}^K a_l(k)  = 0 \\
  D_{l}\left(\Psi\left\{\mathcal{H} \cup l \right\}   \right)   & \text{ otherwise }        
\end{array} \right.
\end{equation} 
where  $\mathcal{H}$ is the  set of the DUs already transmitted in the time slots before $t$ (i.e., $\mathcal{H}$    represents   the scheduling history),   and $D_l$ is the overall distortion level derived from Eq.~\eqref{eq:RD_2}, where the subscripts $\{t,m\}$ have been replaced by the subscript $l$ to describe the data unit $l$. 
%
%$D_0$ is the distortion level of the DU reconstructed only  from the DUs previously transmitted (i.e., the DUs in  $\mathcal{H}$), and $\Delta  D_l$ is the distortion gain due to  the scheduling policy $\pi_l$. 
%The distortion level $D_0$ and the distortion gain $\Delta D_l$ are derived from Eq.~\eqref{eq:RD_2}, where the subscripts $\{t,m\}$ have been replaced by the subscript $l$ to describe the data unit $l$. In particular, 
%\begin{align}
%D_0(\mathcal{H}) & = D_{l}\left( \Psi\left\{\mathcal{H}\right\}  \right)  \nonumber \\
%\Delta D_l(\mathcal{H})  & =  D_{l}\left( \Psi\left\{\mathcal{H}\right\}  \right)   -  D_{l}\left(\Psi\left\{\mathcal{H} \cup l \right\}   \right) , \text{ if }  %\sum_{k=1}^K a_l(k)  >0  
%\end{align}
%$0$ otherwise, where  
The function  $\Psi\left\{\mathcal{H}\right\} $    evaluates the received rate vector $\mat{R}$ of the $M$ views acquired in the last $\rhot$ instants  given the set of transmitted DUs $\mathcal{H}$.  In particular, each element $j$ of the vector $\mat{R}$  is set to $B_j$ if the $j\in\mathcal{H}$, and to $0$ otherwise.  
The evaluation of $D_l$ obviously involves  the size and the prediction maps of the data unit, namely $B_l$ and $\{\phi_{j,l}\}$. For the sake of clarity, we omit this dependency in our equations.

We now consider the rate and distortion for  multiple DUs.
In the joint scheduling of  multiple \acp{DU}, we evaluate    the  average distortion and rate   for a set of scheduling policies  $\mat{\pi}=[{\pi}_1 \ldots {\pi}_L]^T$. This outlines the   dependency between DUs in the packet scheduling optimization. 
The average rate for a set of $L$ DUs with a transmission policy  $\mat{\pi}$ is thus given by
\begin{align}\label{eq:prob2}
   {\mathcal{R}}(\mat{\pi}) & =   \sum_{l} {\mathcal{R}}\left({\pi_l} \right)  = \sum_{l}  B_l  \left[ \sum_{k=1}^K a_l(k) \right] .
\end{align} 
The  derivation of the average distortion    is not as straightforward as the one of the average rate. In particular, the   rate of a given DU only depends  on the scheduling policy for that DU, while the distortion for a given DU depends on the scheduling policy of the correlated DUs\begin{align}\label{eq:prob1}
  {\mathcal{D}}\left(\mat{\pi},    \mathcal{H} \right)
 & =     \sum_{l=1}^L  \frac{1}{w_l}  D_{l}\left( \Psi\left\{\mathcal{H} \cup \mathcal{P}_{\mat{\pi}}\right\}  \right)    
\end{align} 
%\begin{align}\label{eq:prob1}
  %{\mathcal{D}}\left(\mat{\pi},    \mathcal{H} \right)
 %& =  D_0\left(\mathcal{H} \right) - \frac{1}{w_l}  \sum_{l=1}^L   \Delta D_l \left(\mat{\pi},  \mathcal{H}\right)  
%\end{align} 
where  $D_l$  is the   distortion   for the  reconstructed   DU $l$, given the scheduling policy $\mat{\pi}$, and
  $\mathcal{P}_{\mat{\pi}}$  is the set of  DUs scheduled  in the time slots $[t,t+K-1]$ based on the scheduling policy $\mat{\pi}$. Note that, among the DUs in   $\mathcal{P}_{\mat{\pi}}$,     the frames correlated with the DU $l$ have  an impact in the reconstruction of the $l$-th DU in the case where  it cannot be transmitted (i.e., in the case $l \notin \mathcal{P}_{\mat{\pi}}$). 
  
%The value   $D_0$   is the averaged distortion level if none of the $L$ DUs are   transmitted. This means that $D_0$ expresses the distortion when the considered DUs are reconstructed from the frames that have been previously received. It follows that 
%\begin{align}
%D_0(\mathcal{H}) & = \frac{1}{w_l} \sum_{l=1}^L D_{l}\left( \Psi\left\{\mathcal{H}\right\}  \right)   .
%\end{align}
%The evaluation of the distortion gain $\Delta D_l$ is a function of the scheduling policy for all DU's in the scheduling window that are  correlated with the $l$-th  DU. In particular, in the set  $\mathcal{P}_{\mat{\pi}}$, which is the set of  DUs scheduled  in the time slots $[t,t+K-1]$ based on the scheduling policy $\mat{\pi}$, 
 %  the frames correlated with the DU $l$ have  an impact in the reconstruction of the $l$-th DU if it is missing. It follows that
%\begin{align}
%\Delta D_l \left(\mat{\pi},  \mathcal{H}\right)  & = D_{l}\left( \Psi\left\{\mathcal{H}\right\}  \right)      - D_{l}\left( \Psi\left\{\mathcal{H} \cup \mathcal{P}_{\mat{\pi}}\right\}  \right)      
%\end{align}
 Equipped with the above definitions of rate and distortion for each policy, we want now  to find the best scheduling policy $\mat{\pi}^{\star}$  that minimizes the average distortion while satisfying the bandwidth constraints. In particular, we seek for 
\begin{equation}\label{eq:prob}
\mat{\pi}^{\star}(\mathcal{H}) =  \text{arg }  \min_{\mat{\pi}}   {\mathcal{D}}(\mat{\pi},    \mathcal{H}) \text{  \  s.t.  \ }   {\mathcal{R}}(\mat{\pi}) \leq C_{\text{BW}}^{\star}
\end{equation}
where $C_{\text{BW}}^{\star}$ is the   bandwidth constraint given by $C \cdot K \cdot T_{TDMA}$, where $C$ is the channel capacity and $T_{TDMA}$ is the TDMA slot duration in terms of seconds.  In the following, we assume $C_{\text{BW}}^{\star}$ to be constant over time. However,   since our scheduling optimization is refined at every scheduling opportunity, the model can be extended to any system where the bandwidth constraint evolves in time simply by changing the constraint in Eq. \eqref{eq:prob}. 

 Due to the dependency among DUs in Eq.~\eqref{eq:prob1}, the optimization problem  can unfortunately not  be  decomposed easily into mutually independent subproblems. 
 The optimization problem  can be solved with  exhaustive search methods, which however rapidly become computationally intractable for  a large time horizon $K$ and a large number of cameras $M$. An alternative solution consists in solving  the optimization problem with iterative algorithms,  where policies are optimized sequentially.
  The  authors  in~\cite{ChoMia:J06}, for example, propose  an \ac{ISA} method  where, at each iteration, the transmission policy of a single DU is optimized, keeping the other   policies fixed. The overall process is then repeated till convergence.   Unfortunately, due to multiple dependencies between DUs in our problem,   the iterative method  does not necessarily reduce  the computational complexity compared to an exhaustive search strategy.  In the following section, we describe our approximate yet effective solution to determine the best packet scheduling over the time horizon  of size $K$.

%%%%%%%%%%%%%%%%%%%%%%%%%%%%%%%%
\section{Trellis-Based Optimization Solution}
\label{sec:trellis}
We   propose     in this section  a  new trellis-based   method for determining the packet scheduling policies.  The key idea to limit the computational complexity relies on an effective  pruning strategy   based on correlation information. We build a trellis in the solution space as follows.   We consider the scheduling optimization problem over the time horizon  $[t,t+K-1]$.  In the following, we refer to the time instant $(t+k-1)$ as the $k$-th scheduling opportunity (or time slot),  with $k\in[1,K]$. 
  At the $k$-th scheduling  opportunity,  the $L$ DUs that are candidates for scheduling are represented by the states (or nodes) $\{S_{k,1}, \ldots, S_{k,L}\}$. Then, a direct edge (or branch) from state $S_{k,j}$ to the state $S_{k+1,i}$ represents the decision of  scheduling the $i$-th DU  at the $(k+1)$-th transmission opportunity, given that the $j$-th DU  has been transmitted during the $k$-th slot\footnote{From here onwards, ``branch" or ``DU" will be used interchangeably, assuming that each branch represents a scheduled DU.}. A cost   $B_i$ is associated to such an edge, which corresponds to the size of the $i$-th DU.
   For the sake of completeness, we also consider, for each time slot, the null state $S_{k,0}$. A branch  heading  to  the null state means that no frame  is scheduled, and a zero  transmitting rate is associated to this edge. 
A sequence of branches  forms a \textit{path} and all possible paths form a \textit{trellis}.   A \textit{full path} is a path connecting a node at the time slot $k=1$ to a node at the time slot $k=K$. It represents a feasible scheduling policy optimized over a time horizon   $K$ as long as the bandwidth constraints are satisfied (i.e., the sum of the sizes of all transmitted DUs is smaller than the channel capacity). The feasible policy with  the minimum distortion  is   the one leading to the best scheduling policy.  Note that, since we do not consider packet retransmissions in our system,    the  transmission state can only appear once on a path for a given packet.  
 \begin{figure}[t]
\begin{center}
 {\subfigure[ ]{
\includegraphics[width=0.85\linewidth,  draft=false]{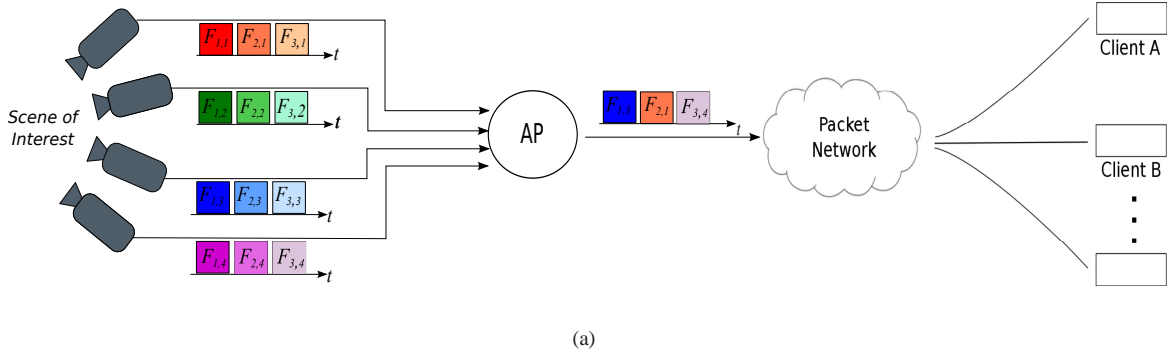}
\label{fig:Trellis_Scenario} }}  \hfill   
\end{center}
\begin{center}
{ \subfigure[]{
\includegraphics[width=0.75\linewidth,  draft=false]{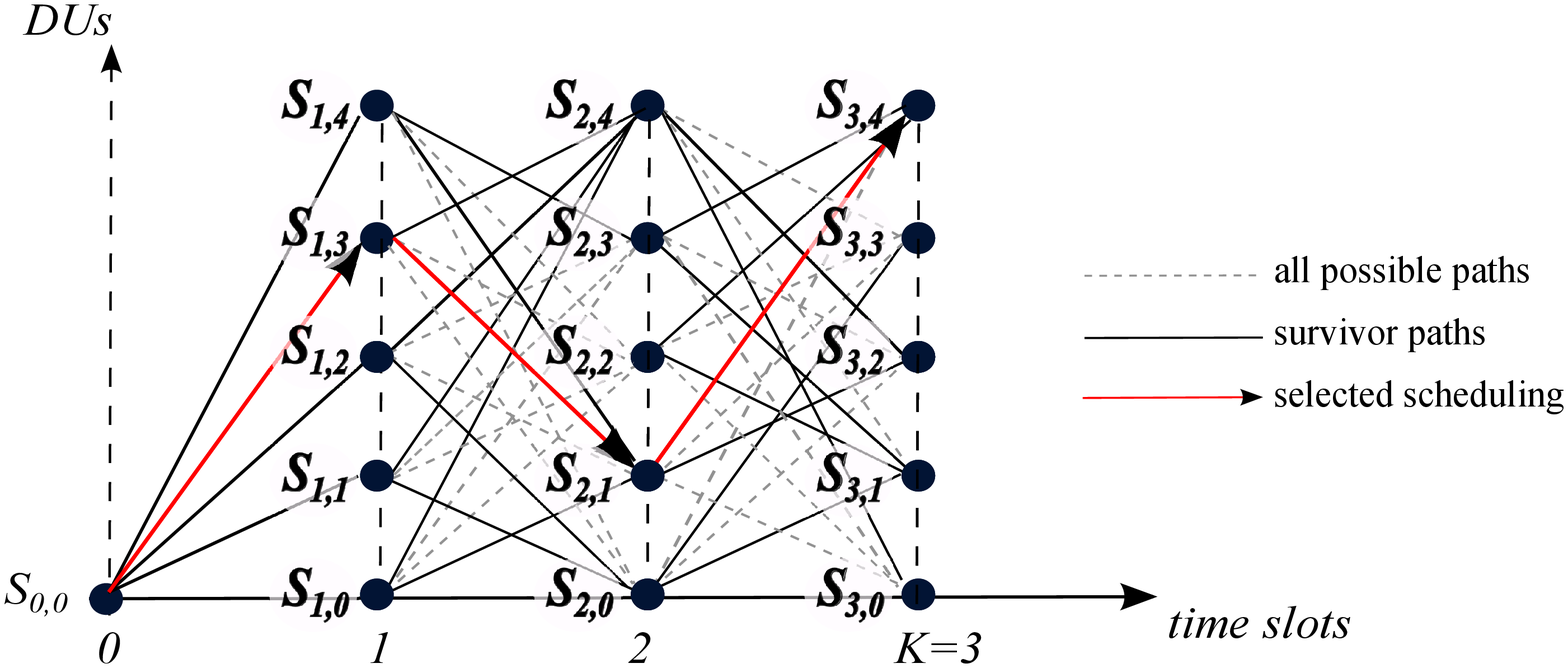}
\label{fig:Trellis_constr} }}   \caption{Example of scheduling policy in a scenario of $4$ cameras and $K=3$ transmission time slots  (a) and  its associated path in the trellis (b).}\label{fig:Trellis}
\end{center}
\end{figure}

 An example of the trellis-based representation  is depicted in Fig.~\ref{fig:Trellis}, where the scheduling policy considers  a time horizon of $K=3$ in a scenario with four cameras. Before starting the frame transmission (i.e., at the time slot $0$) no DUs have been acquired and only the null state is available.  In the general case,  a scheduling policy at the first time slot (i.e., $k=1$)  is represented by a branch going from a specific state $S_{0,i}$ to any possible  state $S_{1,j}$, where   $S_{0,i}$  is the state associated to the DU previously scheduled at the time slot $(t-1)$.   The selected scheduling policy is the one that allocates $F_{1,3}$, then   $F_{2,1}$, and finally $F_{3,4}$.

 As already mentioned above, while the transmitted rate associated to each branch does not depend on the other branches,   the average distortion ${\mathcal{D}}(\mat{\pi}^{\star})$  cannot be evaluated separately for each data unit. Because of the correlation   between DUs,  the distortion   of a given full path is not equal to the summation of the distortion gain for each branch on the path. From an algorithmic point of view, this means that all the branches have to be considered for computing the optimal scheduling solution. Ideally, an exhaustive search should evaluate distortion on all full paths to select the policy with minimum distortion. However, the number of states and full paths are prohibitively large. For example, in a scenario in which $L$ DUs can be scheduled over $K$ time slots, the number of possible full paths is at least  $L! / (L-K-1)!$. 
Rather than an exhaustive search,    we propose a suboptimal algorithm that reduces the visited states 
per time slot and thus substantially reduces  the number of full paths to be tested. The key concept is that the best scheduling policy is likely to be the policy that permits the reconstruction of   most of the scene. Hence the scheduler shall try to send as much ``innovation"  as possible, or as little redundancy as possible. Intuitively, once  a DU is transmitted, the other DUs that carry correlated information should get a smaller priority. The corresponding branches in the trellis are thus unlikely to be part of the optimal path.  
Thus, we propose to prune branches depending  on the level of correlation that exists  between a DU that is candidate for transmission and  the set of  previously scheduled DUs, denoted by ${\mathcal{P}}_{\mat{\pi}^{k}}$ where $\mat{\pi}^{k}$ is the scheduling choices (or path) from $1$ to $k$.

In more details,  we  introduce a  branch reward parameter for each branch in the trellis.  It is  an estimate of the   contribution that the   DU associated to a given branch  can provide to   the overall scene reconstruction process, conditioned on the data that have already been scheduled.   
  Consider a given path ${\mat{\pi}}^{k}$   as the set of DUs scheduled at the first $k$ scheduling opportunities.  We evaluate the   gain of adding  an edge    reaching the node $S_{k+1,q}$ to the path ${\mat{\pi}}{k}$:  we are interested in the reward of scheduling  the DU $q$   at the time slot $k+1$, given that the DUs in the set ${\mathcal{P}}_{{\mat{\pi}}^{k}}$   have  been previously scheduled.  This branch reward is formally  given by
\begin{align}\label{eq:profit}
\rho(S_{k+1,q} | {\mathcal{P}}_{\mat{\pi}^{k}} ) = \frac{1}{L} \sum_{l=1}^L \left\{  \sum_{s_j\in F_l }\alpha(s_j)  \max\left\{ 0, \left[ \mat{\phi}_{j,l} \cdot \Psi\left\{{\mathcal{P}}_{\mat{\pi}^{k}} \cup q\right\} -   \mat{\phi}_{j,l} \cdot \Psi\left\{{\mathcal{P}}_{\mat{\pi}^{k}} \right\}\right]  \right\}  \right\}
\end{align}
 In other words, the reward   $\rho(S_{k+1,q} | {\mathcal{P}}_{\mat{\pi}^{k}} )$ is the ``innovative" contribution that the DU $q$ can offer to the reconstructed scene.  In particular, for the decoding of the $l$-th DU among the $L$ DUs under consideration, $\max\left\{ 0, \left[ \mat{\phi}_{j,l} \cdot \Psi\left\{{\mathcal{P}}_{\mat{\pi}^{k}} \cup q\right\} -   \mat{\phi}_{j,l} \cdot \Psi\left\{{\mathcal{P}}_{\mat{\pi}^{k}} \right\}\right]  \right\} $ is equal to $0$ if the region $s_j\in F_l$ can be reconstructed from the previously scheduled  DUs (i.e., the DUs in ${\mathcal{P}}_{\mat{\pi}^{k}}$), while it is equal to $1$ if the region  cannot be reconstructed from the DUs in  
${\mathcal{P}}_{\mat{\pi}^{k}} $. In the latter case, the DU $q$ is innovative for the region  $s_j$.

% \begin{figure}[t]
%\begin{center}
%\includegraphics[width=0.75\linewidth,  draft=false]{figure/Trellis_SubOptimal_3.eps}
%\caption{Example of suboptimal trellis-based optimization when $N_s=2$.}
%\label{fig:Trellis_SubOptimal}%
%\end{center}
%\end{figure}

 We now describe our solution    to optimize the   scheduling policy  at time $t$ and over a time-horizon of $K$; the key concept is that, at each scheduling opportunity, we select a subset of all branches defined in the trellis and we consider the subset as the search  space for our packet scheduling policy.  The   branches in the subset are selected  as the ones with the highest branch reward in Eq. \eqref{eq:profit}.   We assume that, at time $t$ (i.e.,  $k=1$),   all branches represent possible candidates for being the first part of the best scheduling solution (i.e., no pruning is done on the first branch of the paths).   
   Thus,    we initially determine $\{\mat{\pi}^1\}$ as the set of branches going from the time slot $k=0$ (i.e., the node representing the scheduling history) to the time slot $k=1$.  In general, we denote by $\{\mat{\pi}^k\}$ the set of all paths from $1$ to $k$  (i.e., the set of possible scheduling policies in the first $k$ time slots), and by $\mat{\pi}^k$ a generic element of the set.
For each path $\mat{\pi}^k$,  the search space of     possible branches in which the current path can be extended is denoted by ${\mathcal{B}}_{\mat{\pi}^{k}}$. From ${\mathcal{B}}_{\mat{\pi}^{k}}$, a subset of at most $N_s$ survivor branches are selected as the ones  satisfying the bandwidth constraints and  maximizing the branch profit $\rho(S_{k+1,q} | {\mathcal{P}}_{\mat{\pi}^{k}} )$, with $q\in{\mathcal{B}}_{\mat{\pi}^{k}}$. 
This means that  $N_s$ branches will be considered for  constructing the  candidate paths $\mat{\pi}^{k+1}$  starting from  $\mat{\pi}^k$.
This subset selection is evaluated for each element in $\{\mat{\pi}^k\}$ and successively  for all the  $k>1$. This leads to at most $N_s^{K-1}$  possible  paths for each  $\mat{\pi}^1$. 
 Once   the full paths  are evaluated, we identify the best scheduling policy as the one that corresponds to the full path minimizing the overall distortion. 
  The overall scheduling algorithm is presented in Algorithm~\ref{alg1}.  The branch pruning strategy  allows us to explore only   $\left(|\{\mat{\pi}^1\}|  N_s^{K-1}\right)$ paths at most.

\begin{algorithm}[t]                      % enter the algorithm environment
\caption{Scheduling  Optimization Algorithm}          % give the algorithm a caption
\label{alg1}
\begin{algorithmic}[1]                    % enter the algorithmic environment
\vspace{.3cm}
\REQUIRE \textnormal{Set $k=0$. Select all possible branches from the single state in $k=0$ to all defined states in $k=1$. Denote by $\{\mat{\pi}^1\}$ the set of all branches from $k=0$ to $k=1$, and by $\mat{\pi}^1$ a generic element of the set. }  
\FOR{$k=1$ to $K-1$}
\FOR {\textnormal{each path} $\mat{\pi}k \in \{\mat{\pi}^k\}$}
\STATE \textnormal{\textit{step a)}: for the considered path from $0$ to $k$, individuate all branches going from the scheduling opportunity $k$ to the scheduling opportunity $k+1$. Denote by ${{\mathcal{B}}}_{\mat{\pi}k }$ the set of these  branches. }  
\STATE \textnormal{\textit{step b)}: among  branches in  $ {{\mathcal{B}}}_{\mat{\pi}^k}$ that satisfy  the bandwidth constraints  identify the subset of the  $N_s$    branches with the highest profit $\rho(S_{k+1,q} | {\mathcal{P}}_{\mat{\pi}^{k}} )$, with $q \in  {{\mathcal{B}}}_{\mat{\pi}^k}$ and discard the remaining  branches.  }
\STATE \textnormal{\textit{step c)}: include the $N_s$ selected paths (i.e., the considered path $\mat{\pi}^k$ plus the $N_s$ selected branches) in $\{\mat{\pi}^{k+1}\}$.}
\ENDFOR
\STATE $k \leftarrow k +1$.
\ENDFOR
\STATE \textnormal{evaluate the best scheduling policy  $\mat{\pi}^{\star}$ as        $$ \mat{\pi}^{\star}=\arg \min_{\mat{\pi}  \in \{\mat{\pi}^K\}} {\mathcal{D}}(\mat{\pi}) \text{ s.t. } {\mathcal{R}}(\mat{\pi}) \leq C_{\text{BW}}^{\star}$$  }
 \end{algorithmic}
\end{algorithm}

An example of our algorithm is depicted in Fig.~\ref{fig:Trellis_constr} for a scenario of $4$ cameras. In this example, for the sake of simplicity, we assume that the decoding deadline is $T_D=1$ such that  each frame acquired at the time slot $k$ expires at the time slot $k+1$.
 We consider the first frame of the sequence and $S_{0,0}$ is the initial state of the scheduler ($t=1$).
 No branch is pruned in  the first time slot. This means that $\{\mat{\pi}^1\}=\{(S_{0,0}-S_{1,0}), (S_{0,0}-S_{1,1}), (S_{0,0}-S_{1,2}), (S_{0,0}-S_{1,3}),(S_{0,0}-S_{1,4}) \}$, where $(S_{q}-S_{q^{\prime}})$ represents the branch going from state $S_q$ to state $S_{q^{\prime}}$. For each of these branches, we evaluate the full paths as follows. Considering $\mat{\pi}^{1}=(S_{0,0}-S_{1,1})$ and $N_s=2$, the subset of survivor branches for $k=2$   is $\{ (S_{1,1}-S_{2,4}), (S_{1,1}-S_{2,0})\}$. These two survivor branches are included in $\{\mat{\pi}^2\}$, and the operation is repeated    for every  branch in $   \{\mat{\pi}^1\}$.   The branch pruning strategy is considered also for $k=3$, obtaining then the set $\{\mat{\pi}^3\}$, which is the set of all the survivor  full paths going from $k=0$ to $k=3$. In our illustrative example, these paths are represented by solid black lines.  
 Among the candidates full paths, we finally select the best  scheduling solution as the one minimizing the distortion as evaluated in Eq.~\eqref{eq:prob}.

%%%%%%%%%%%%%%%%%%%%%%%%%%%
 \section{Simulation Results}
\label{sec:results} 

\subsection{Simulation Setup}
We provide now simulation results for a multi-camera scenario where data have to be transmitted over a bottleneck channel of rate $C_{BW}$.  We start the scheduling optimization at $t=1$.   Since each scheduled DU is entirely transmitted, we consider the next transmission opportunity as $t+T_u$, where $T_u$ is the number of time slots required to transmit the selected DU.  At this new scheduling opportunity, a new optimization is performed over the successive $K$ time slots. We proceed similarly till the end of the simulation, which in our case corresponds to the expiration time   of the last frame of the video sequence.  
 
 We consider image sequences where all the DUs from all the cameras have the same size $R$ for the sake of simplicity, and assume that all the views have the same importance, i.e., $w_m = w$ in Eq.~\eqref{eq:RD_1}. Our simulations are carried out with the ``Ballet" and ``Breakdancer" video sequences \cite{web_microsoft_ballet_break}, which consist of $N_f = 100$ frames, at a resolution of $S_R = 768 \times 1024$ pixel/frame and $F_R=15$ frames per second. The total number of camera views ranges from 4 to 8. We study the performance of our algorithms in different configurations, for different camera setups, different  values of the DU size $R$ and for different constraints on the bottleneck bandwidth  $C_{BW}$. 

We denote by $\rhos$ the number of  spatially correlated cameras and we  assume that each view is correlated to $\rhos/2$ neighbor 
views,   if available, on both the left and the right sides.   
As already mentioned in Sec. \ref{subsec:RD}, the correlation in time, denoted by $\rhot$, is related to the   number of frames considered in temporal interpolation  at  the  decoder.
Both $\rhot$ and $\rhos$ represent the \textit{maximum} number of correlated frames in the time and space domain, respectively. The \textit{actual} level of correlation experienced in each single frame depends also on the video content. The control parameters $\rhot$  and $\rhos$  take different values  in our simulations in order to study the behavior of the scheduler for   {different correlation image reconstruction scenarios.  }
We experimentally build the $\mat{\phi}$ matrix as explained in Sec. \ref{sec:frameworks}. In the Appendix we provide some further details on the construction of the matrix. 
In short, the number of regions  in which each frame is subdivided depends on both the video content and   the correlation level. Thus,   frames can be decomposed into different regions. In particular, each region is designed by  a unique combination of correlated frames that are involved in the reconstruction at the decoder. 
In the temporal domain,   the contribution of neighboring frames to each region is evaluated by comparing images from the same camera. 
More precisely, each frame is subdivided into regions, each of them can be reconstructed from     previously  acquired frames only if no  motion occurs in these regions. As   no motion estimation  is employed at the source coding nor at the receiver in our system,   only the fixed background contributes to the   temporal extrapolation of missing frames.   
 In the spatial domain,  to evaluate the influence of each camera on the  neighboring ones,  
  we use DIBR techniques and calculate the number of pixels that can be estimated from neighboring views. This can be achieved by each camera with the information about its own depth map, and about the positions of the neighbor cameras.  The overhead information required for this estimation thus corresponds to the information about  the camera positions, which is generally of small size.  As observed in  \cite{TonMauFro:C12}, the exact value of the correlation level is however not a critical parameter in the scheduling optimization. Errors in the correlation evaluation, caused by a   coarser estimation with a smaller  overhead, does not have a significant impact on the scheduling policies. Thus, in the following, we assume a precise knowledge of the correlation information and we neglect the small overhead required to estimate the correlation level.

Since we are interested in reconstructing all the views   (at the clients), simulation
results are provided in terms of mean PSNR, which is the PSNR averaged over all the frames of all views.
This means that, even if some frames are decoded at high PSNR values, the average  PSNR of the reconstructed scene might be in the low PSNR range in challenging transmission conditions.   First, the PSNR of the reconstructed scene  is evaluated from the rate-distortion model  described in Sec.~\ref{sec:RD}. Then we validate our findings by experiments with actual reconstruction of the video frames at the decoder.  
 
 The proposed algorithm has been compared to two baseline algorithms: a random allocation  of the DUs (``Baseline - RNDM"), whose distortion performance has been averaged   over $1000$ runs, and a scheduling solution where cameras priorities are defined a priori based on the joint entropy of the camera dataset as defined in \cite{DaiAky:J09} (``Baseline -
Akyildiz"). In particular, the camera selection for the latter method is  based on the spatial correlation that exists between views, while  time correlation information is neglected. The camera priority is established as follows: the camera   minimizing the overall distortion becomes the highest priority camera.  Then,   other cameras are  included if they maximize  the diversity (i.e., if they minimize  the spatial correlation) with respect to  the cameras  that have been previously selected.  
We first  provide results for a greedy optimization scenario (i.e., $K=1$) and demonstrate the benefit of a correlation-aware scheduling optimization w.r.t. baseline algorithms. Then we depict the performance of foresighted optimization solutions, showing that    low-complexity solutions lead to good performance when the     optimization horizon is enlarged.

\subsection{Greedy Optimization}
We first analyze the performance of our algorithm in the case where the optimization horizon is limited to the next transmission time slot. 
We first study the importance of the knowledge of the correlation information in the optimization.
Our optimization algorithm is evaluated in different conditions that depend  on the type of correlation information   considered in the scheduling decisions:
i) ``Correlation Known", when the full correlation information is considered in the optimization;
ii) ``Space Corr Known", when only the spatial correlation is considered;
iii) ``Time Corr Known", when only the temporal correlation is used;
iv) ``No corr known", when the scheduler completely ignores the correlation between frames.

 \begin{figure}[t]
\begin{center}
\includegraphics[width=3.1in,draft=false]{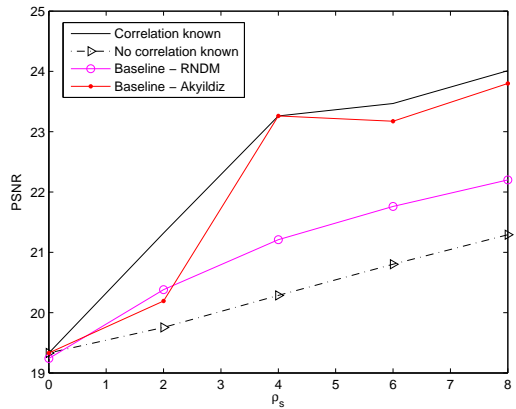}
\caption{PSNR vs spatial correlation level $\rhos$ for systems with $8$
cameras  $(C=23.5\, Mbps$, $ r=11.7\, Mbps$, $\TD=5$,
 and $\rhot=0$, Ballet sequence model).  }
\label{fig:Ballet_perTconc_N8_AP5_Tacq4_Rate10_Tc0}
\end{center}
\end{figure}

%%%%%%%%%%%%%%%%%%%%%%%%
%%%% BREAKDANCER %%%%%%%%%%%
% \begin{figure}[t]
%\begin{center}
%\includegraphics[width=3.5in,draft=false]{figure/FIG6_B_Breakdancer_PSNRvsSpatCorr_perTconc_N8_AP5_Tacq4_LF0_DF0_Tc0_Rate100_Ttdma50.eps}
%\caption{PSNR vs spatial correlation level $\rhos$ for systems with $8$
%cameras  $(C=23.5\, Mbps$, $ r=11.7\, Mbps$, $\TD=5$,
% and $\rhot=0$, Breakdancer sequence model).  }
%\label{fig:Breakdancer_perTconc_N8_AP5_Tacq4_Rate10_Tc0}
%\end{center}
%\end{figure}
%%%%%%%%%%%%%%%%%%%%%%%%

We first study the gain that can be achieved when
the correlation model is known by the scheduler.   In the following figures,
the PSNR of the reconstructed scene  is evaluated from the rate-distortion model  described in Sec.~\ref{sec:RD}. 
In the first experiments reported in Fig.  \ref{fig:Ballet_perTconc_N8_AP5_Tacq4_Rate10_Tc0},   the  temporal correlation between cameras is neglected both at the scheduler and at the decoder and we focus on the influence of the spatial correlation, which means that missing frames are reconstructed from neighboring views but not from previous frames. 
 The performance of the scheduling algorithm is given as a
function of the spatial correlation $\rhos$ (i.e., a function of the number of views that are considered to be spatially correlated) for systems with $8$
cameras, a playback delay $\TD=5$,  a constant  encoding rate per camera  of $r=11.7\, Mbps$ and a channel capacity  $C=23.5\, Mbps$\footnote{Note that $r = R[bpp] \cdot S_R[\text{pixel per frame}] \cdot F_R[fps].$}. 
This  bandwidth constraint means   that $2$ only frames out of $8$ can be allocated on the channel between each frame acquisition.
 First, we observe that the gain experienced by the algorithm using the spatial correlation information in the scheduling compared
to the case in which all the correlation levels are ignored is
substantial and this gain increases with the number of correlated frames (i.e., with  $\rhos$). Thus, the knowledge of the spatial correlation is able to considerably improve the
efficiency of the scheduling decisions.
Moreover, the proposed
algorithm   outperforms both
  baseline algorithms.
This means that  the packet scheduling optimization leads to a better
level of adaptation than the a priori camera selection technique in~\cite{DaiAky:J09}. It is interesting to note  that, by neglecting the correlation model (``No Correlation Known") the performance becomes very bad and   even worse than  a random allocation solution.    This means that, rather than choosing the scheduling based  on   wrong correlation information, it is better to completely ignore it.  
%This is due to the fact that a random allocation might select either a very good   policy or a very bad policy. On the contrary, a scheduling optimization that neglects the correlation model, \new{i.e.,  that relies on a model different from the one used in the reality, most likely takes always wrong decision, leading to a   bad scheduling policy}.

\begin{table*}[t]
\begin{center}
\caption{Average PSNR  of the reconstructed images for each camera  for systems with $8$ cameras   $(\rhos=8$, $\rhot=3$,  $C=23.5\, Mbps$, $ r=11.7\, Mbps$, and $\TD=5)$, for the Ballet sequence model.} \label{table:single_view}
\begin{tabular}{|c|c|c|c|c|c|c|c|c|} \hline
\multirow{2}{*}{Optimization Method}
 &   \multicolumn{8}{c|}{Camera view}  \\   \cline{2-9}
 & 1& 2& 3& 4& 5& 6& 7& 8   \\ \hline \hline 
No Correlation known &    24.95 &  25.32 & 26.97 &  27.44 & 26.88 & 26.69 & 25.80 &  25.26   \\ \hline
 Correlation known &   26.19 & 26.26 & 24.13 &  28.08  &  26.23 &  25.18 &  26.87 &  26.18 \\ \hline
Baseline - Akyildiz &          22.28   &   23.07 &  24.87 & 24.52 & 24.64 & 25.84 & 23.84 &  22.55  
  \\ \hline   
\end{tabular}
\end{center}
\end{table*}
   \begin{figure}[t]
\begin{center}
\includegraphics[width=3.1in,draft=false]{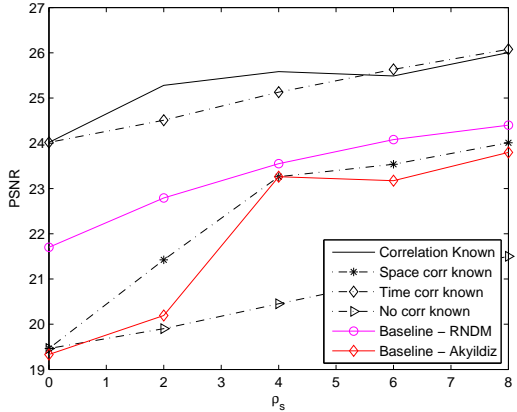}
\caption{PSNR vs spatial correlation level $\rhos$ for systems with $8$ cameras  ($C=23.5\, Mbps$, $ r=11.7\, Mbps$,  $\rhot=3$,
 and $\TD=5$, Ballet sequence model).}
\label{fig:Ballet_perTconc_N8_AP5_Tacq4_Rate10_Tc3}
\end{center}
\end{figure}

%%%%%%%%%%%%%%%%%%%%%%%%
%%%% BREAKDANCER %%%%%%%%%%%
%\begin{table*}[t]
%\begin{center}
%\caption{Average PSNR  of the reconstructed images for each camera  for systems with $8$ cameras   $(\rhos=8$, $\rhot=3$,  $C=23.5\, Mbps$, $ r=11.7\, Mbps$, and $\TD=5)$, for the Breakdancer sequence model.} \label{table:Breakdancer_single_view}
%\begin{tabular}{|c|c|c|c|c|c|c|c|c|} \hline
%\multirow{2}{*}{Optimization Method}
% &   \multicolumn{8}{c|}{Camera view}  \\   \cline{2-9}
% & 1& 2& 3& 4& 5& 6& 7& 8   \\ \hline \hline 
%No Correlation known &   25.63  &  25.29  &    25.58 & 24.89  &  24.71 &     23.80 & 22.83 & 22.15     \\ \hline
% Correlation known &  26.69 & 27.45 & 27.41 & 27.47 & 26.88 &  28.04 &  27.23 &  25.17    \\ \hline
%Baseline - Akyildiz &   25.24 &  26.34 & 28.22 &  24.61 &  25.13 &  27.57 &  26.39 &  25.70 
%  \\ \hline   
%\end{tabular}
%\end{center}
%\end{table*}
%
%   \begin{figure}[t]
%\begin{center}
%\includegraphics[width=3.5in,draft=false]{figure/FIG7_B_Breakdancer_PSNRvsSpatCorr_perTconc_N8_AP5_Tacq4_LF0_DF0_Tc3_Rate100_Ttdma50.eps}
%\caption{PSNR vs spatial correlation level $\rhos$ for systems with $8$ cameras  ($C=23.5\, Mbps$, $ r=11.7\, Mbps$,  $\rhot=3$,
% and $\TD=5$, Breakdancer sequence model).}
%\label{fig:Breakdancer_perTconc_N8_AP5_Tacq4_Rate10_Tc3}
%\end{center}
%\end{figure}
%%%%%%%%%%%%%%%%%%%%%%%%

In the next experiment,    temporal correlation    is considered in the scheduling decisions.   The PSNR quality is provided 
in Fig.~\ref{fig:Ballet_perTconc_N8_AP5_Tacq4_Rate10_Tc3} 
as a function of the number of spatially correlated cameras $\rhos$  for systems with
$8$ cameras, $C=23.5\, Mbps$, $ r=11.7\, Mbps$ and a temporal correlation  $\rhot=3$ (i.e., each frame is considered to be correlated with the three previous frames of the same camera view).
It can be observed that the algorithm using temporal correlation (``Time Corr Known")  is the closest one to the algorithm using all the correlation information (``Corr Known").  It has to be noted  that  all the results provided in the Fig.~\ref{fig:Ballet_perTconc_N8_AP5_Tacq4_Rate10_Tc3}  have been evaluated considering temporal interpolation at the decoder. However, not all the algorithms include this information in the scheduling optimization. For example, the algorithm that only  takes into account   the spatial correlation information  (``Space Corr Known")  is not able to outperform the baseline algorithm with random allocation. This means that, when views are highly correlated in both temporal and spatial domains, a partial information on the correlation does not always lead to a considerable gain in the scheduling optimization.  
 In Table \ref{table:single_view}, the average PSNR for the sequences reconstructed in the different camera views is provided for the same experiment. It can   be observed that most of the reconstructed camera views achieve the highest PSNR with the correlation-aware scheduling algorithm. 

\begin{figure}[t]
\subfigure[ $\rhot=0$,   $\rhos=0$.]{
\includegraphics[width=3.1in,draft=false]{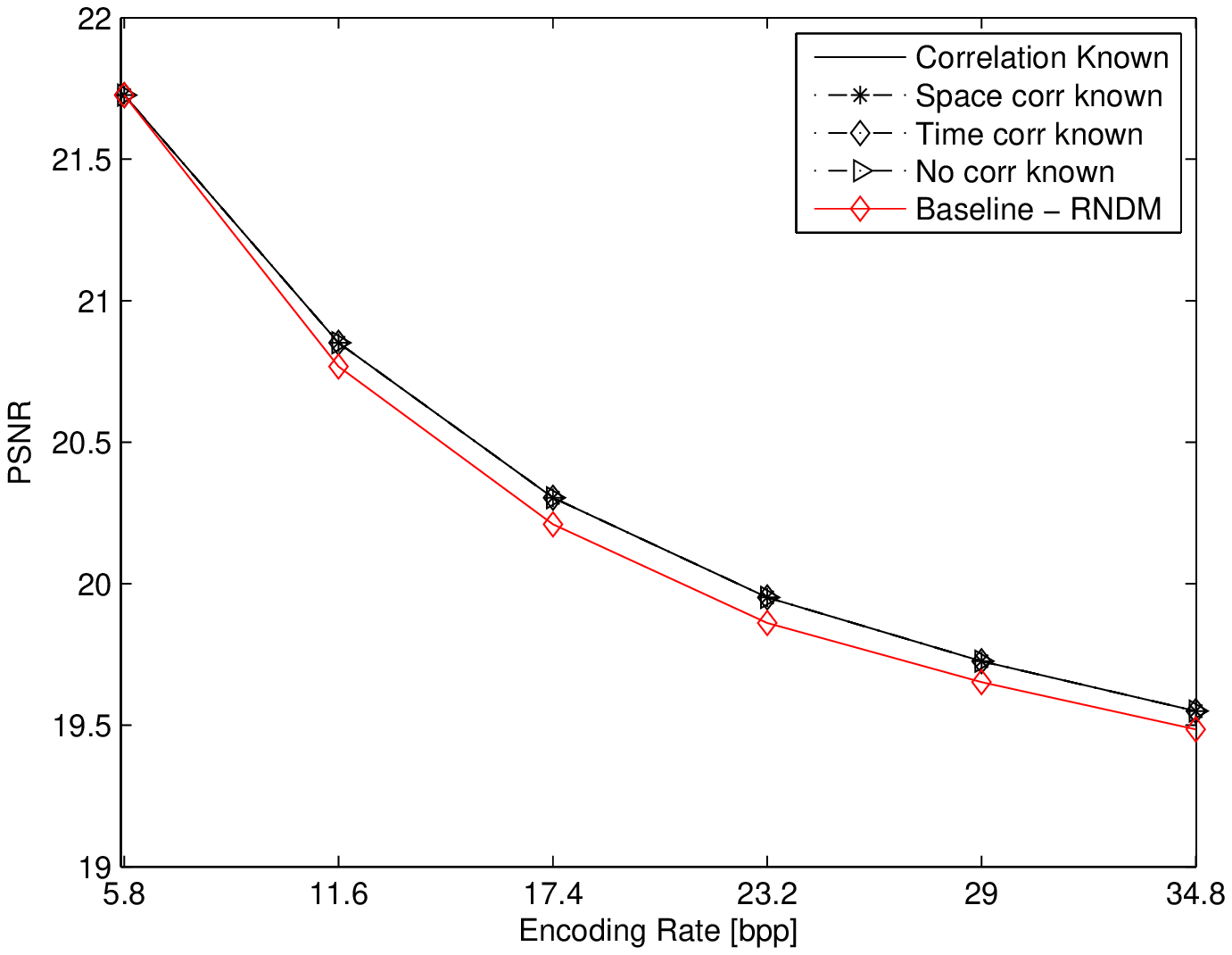}
\label{fig:bitrateA} }
 \subfigure[ $\rhot=2$,   $\rhos=4$.]{
\includegraphics[width=3.1in,draft=false]{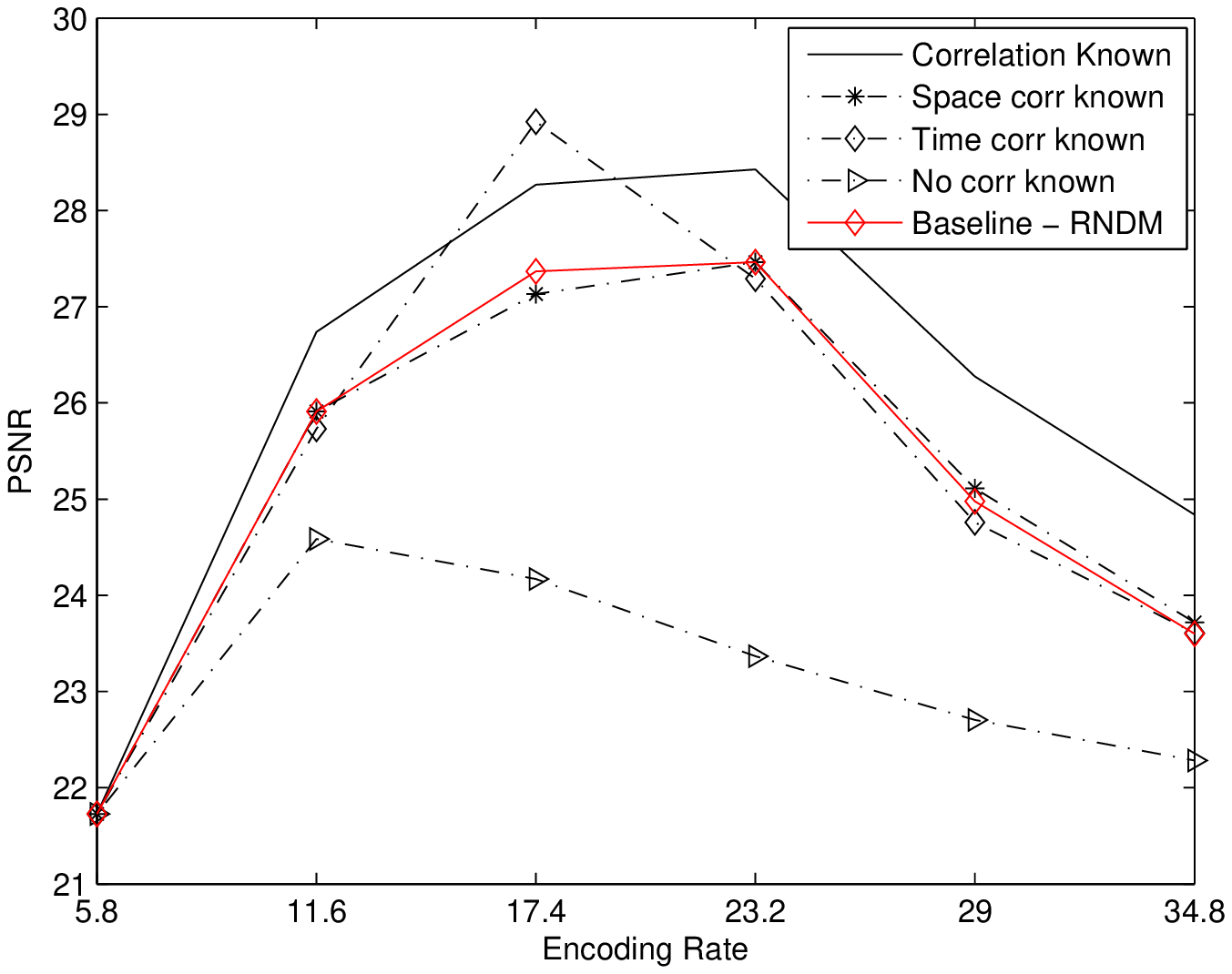}
\label{fig:bitrateB}} \caption{PSNR vs encoding rate for  systems with $4$ cameras  ($C=23.5\, Mbps$, and
$\TD=5$, Ballet sequence model).}
\label{fig:Ballet_PSNRvsBitrate_perTconc_N4_AP5_Tacq4_LF0_DF0_Tc2_SpaceCorrc4}
\end{figure}   

%%%%%%%%%%%%%%%%%%%%%%%%
%%%% BREAKDANCER %%%%%%%%%%%
%\begin{figure}[t]
%\subfigure[ $\rhot=0$,   $\rhos=0$.]{
%\includegraphics[width=3.1in,draft=false]{figure/Ballet_PSNRvsBitrate_perTconc_N4_AP5_Tacq4_LF0_DF0_Tc2_B.eps}
%\label{fig:Breakdancer_bitrateA} }
% \subfigure[ $\rhot=2$,   $\rhos=4$.]{
%\includegraphics[width=3.1in,draft=false]{figure/Ballet_PSNRvsBitrate_perTconc_N4_AP5_Tacq4_LF0_DF0_Tc2_A.eps}
%\label{fig:Breakdancer_bitrateB}} \caption{PSNR vs encoding rate for  systems with $4$ cameras  ($C=23.5\, Mbps$, and
%$\TD=5$, Breakdancer sequence model).}
%\label{fig:Breakdancer_PSNRvsBitrate_perTconc_N4_AP5_Tacq4_LF0_DF0_Tc2_SpaceCorrc4}
%\end{figure}   
%%%%%%%%%%%%%%%%%%%%%%%%

We now repeat similar experiments in a different camera configuration with only 4 views. 
In Fig.~\ref{fig:Ballet_PSNRvsBitrate_perTconc_N4_AP5_Tacq4_LF0_DF0_Tc2_SpaceCorrc4},
the PSNR quality is measured as a function of the encoding rate ($C = 23.5\ Mbps$, $\TD=5$).  It can be observed that there is a tradeoff in the choice of   the   encoding rate, which varies with   the level of   correlation information used in the scheduling decisions. This tradeoff is the result of a source quality that increases with encoding rate, while the penalty due to the channel   also increases with encoding rate, since more DUs are dropped at high rate for the same channel bandwidth constraint. 
If there is no known correlation neither in time nor space (i.e., $\rhos=0$, $\rhot=0$ in Fig.~\ref{fig:bitrateA}),  it is better to reduce the encoding rate,
 so that there is a chance of increasing the number of DUs
allocated for transmission, hence the diversity of the information.
On the contrary, when the correlation can be exploited   both in time and space for
frame interpolation (i.e.,    $\rhos
= 4$, $\rhot = 2$ in Fig.~\ref{fig:bitrateB}), the best encoding rate appears to be a medium rate   ($17\, Mbps$). This means that, in this case, rather than scheduling all the frames at low rate (i.e., $r=5.8\, Mbps$), it is better to transmit less frames but at higher rate  and to   exploit the correlation for the reconstruction of the missing ones.

\begin{figure}[t]
\subfigure[PSNR vs temporal correlation information $\rhot$ for systems with  $4$ cameras  ($C = 23.5\, Mbps$, $r = 23.5\,
Mbps$, $\rhos=2$, and  $\TD=5$).]{
\includegraphics[width=3.1in,draft=false]{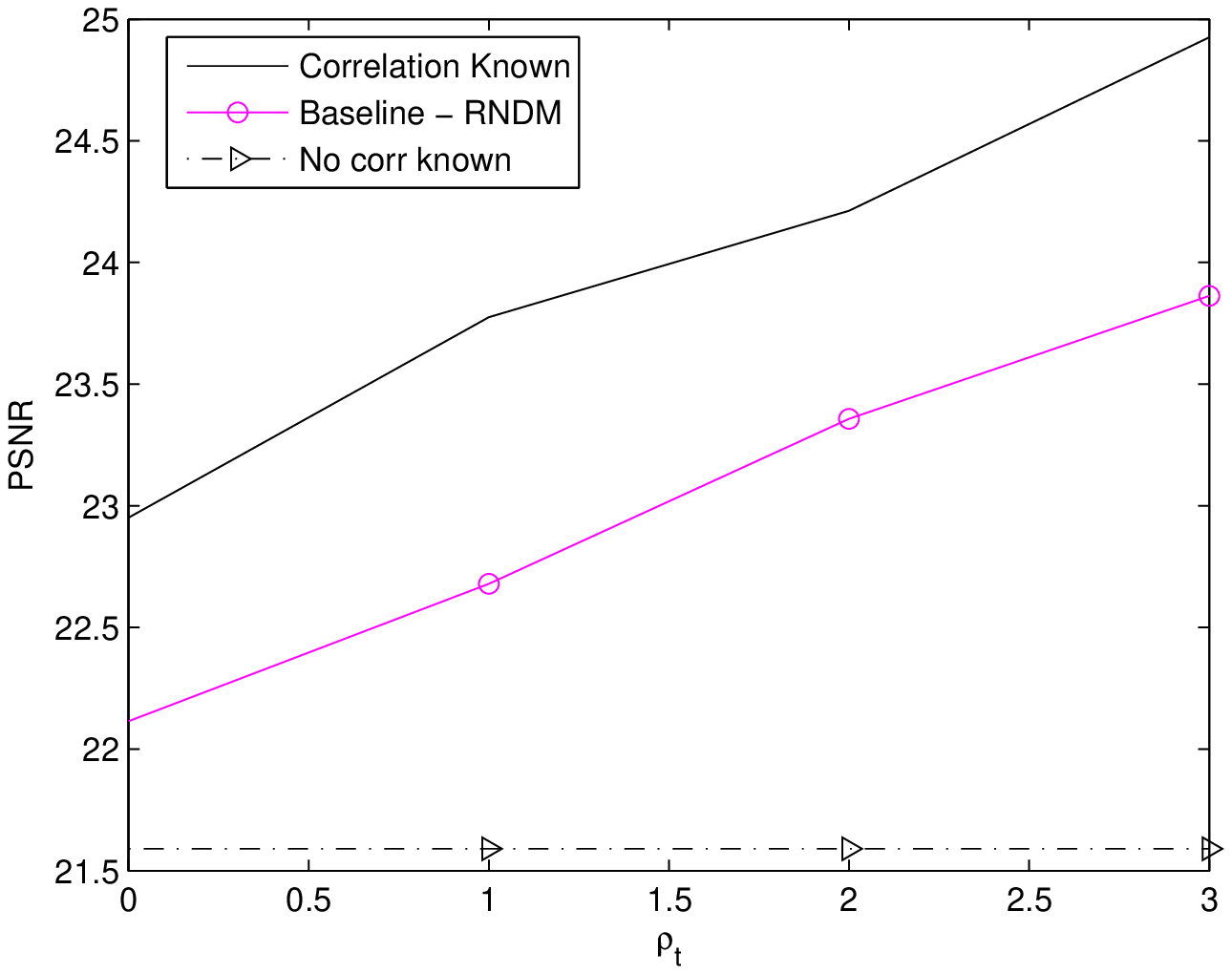}
\label{fig:experimental} }
 \subfigure[PSNR vs spatial correlation level $\rhos$ for systems with  $8$ cameras  ($C = 23.5\, Mbps$, $r = 11.7\,
Mbps$, $\rhot=3$, and  $\TD=5$).]{
\includegraphics[width=3.1in,draft=false]{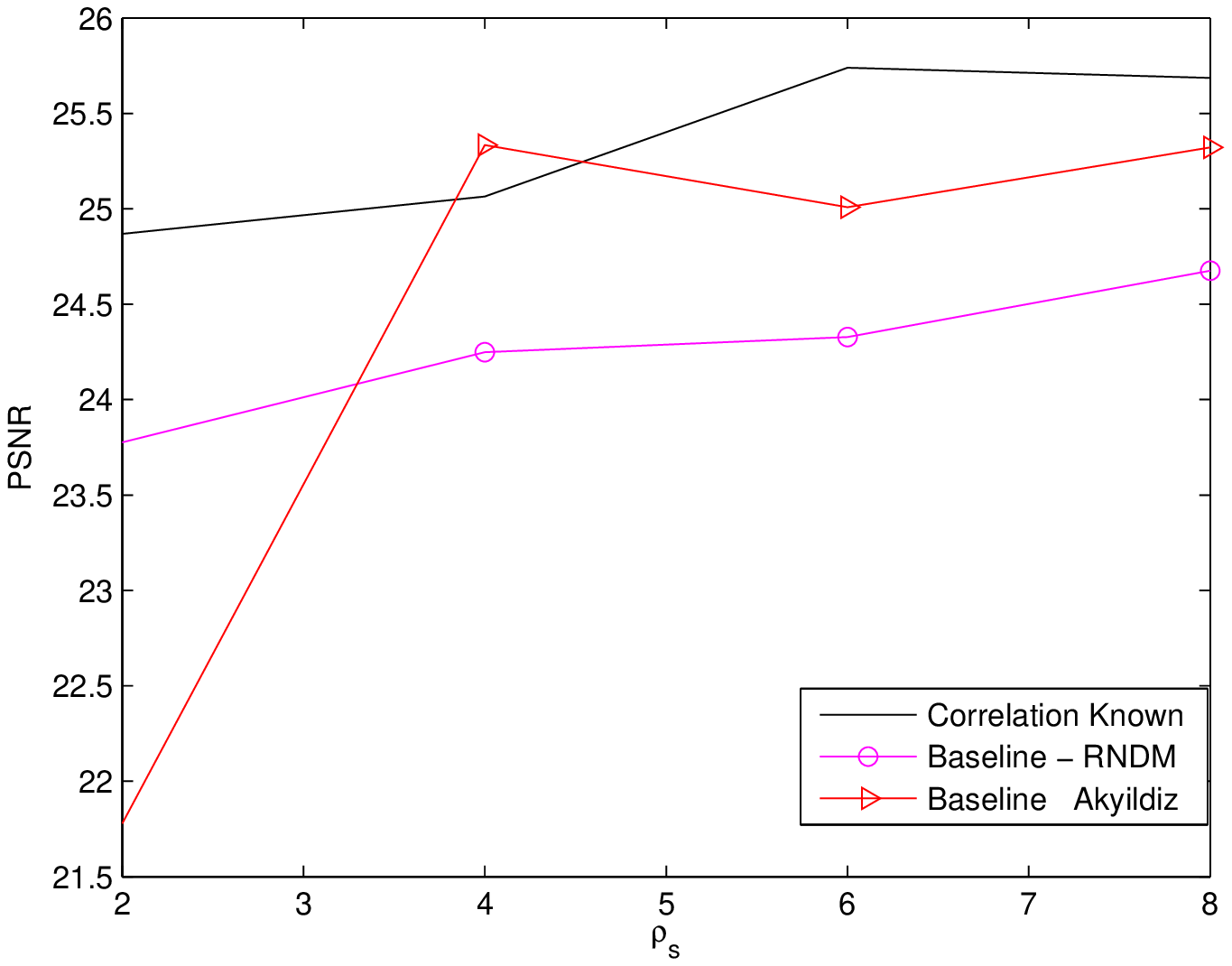}
\label{fig:theory}} \caption{Reconstructed PSNR for systems with $4$ and $8$ cameras for different encoding rates and levels of correlation (Ballet sequence).}
\label{fig:Experimental_Restuls}
\end{figure}

Finally,  we confirm the above observations on experiments with a system that performs actual reconstruction of the video frames at the decoder.  These results  are provided  in  Fig.~\ref{fig:Experimental_Restuls}.
The ``Baseline-Akyildiz" performs better   than a random scheduling   most of the time, 
but it is in general outperformed by the proposed scheduling optimization, for almost all the   values $\rhos$ of spatial correlation.
 These observations are in line with our previous results where the quality is measured with the R-D model of Sec.~\ref{sec:frameworks}. They confirm the benefits of including correlation information in the scheduling algorithm, even in a greedy scenario ($K=1$).
 
%%%%%%%%%%%%%%%%%%%%%%%%%%%%%%%%%%%%%%%%%%%%%%%%%%%%%%%%%%%%%%%%%%%%%
\subsection{ Large Optimization Horizon}
We now provide results for a   framework  with foresighted  optimization where scheduling policies are computed for several future time slots ($K>1$).  We have already  shown above    the gain of the proposed algorithm over the baseline ones from $K=1$, so that we now limit the study to the proposed scheduling algorithm, and look at the gain of a foresighted scheduling policy with respect to a greedy optimization.  
  First, we provide results where  the quality is measured with the R-D model of Sec.~\ref{sec:frameworks} (no actual reconstruction of the video frames at the decoder). Then we validate our findings by experiments with actual reconstruction of the video frames at the decoder.   For the branch pruning strategy in the trellis-based scheduling solution, we consider the number of survivor branches per time slot to be  $N_s=2$.
The results are provided for both a static scenario, where cameras are fixed and the correlation level variations are due to video content,  and a dynamic scenario, where cameras are allowed to move in time with a   dynamic level of   spatial correlation.%, and scenes cut events are also considered by  randomly selecting   time instants in which the acquired frames have no  correlation with the previous ones.  
 The random movement of the cameras is simulated as follows. 
 We assume a set of $2M$ possible positions that each camera can take. We start the simulation  by  randomly allocating each camera in one of the available positions. At each time slot, a camera is randomly selected for changing its position (it can randomly move to the neighboring position).   The camera  moves only if the chosen position is not already occupied by another camera; otherwise no movement is performed by the camera set at this time slot. Based on the position of the cameras, the correlation level is evaluated. This means that
   the correlation between two neighboring cameras can dynamically vary in time, accordingly with the camera movement. In particular, each view can always be reconstructed from the two neighboring ones, but if these two are far apart the portion of frame that can be reconstructed will be small.  Moreover, we also assume that  the correlation with the frame previously acquired in time is zero when there is a camera motion. Each result provided in the following  solution has been averaged over $1000$ simulations runs.  
   
\begin{table*}[t]
\begin{center}
\caption{Average PSNR  of the reconstructed sequence for each camera  for systems with $4$ cameras   $(C=47\, Mbps$, $ r=23.5\, Mbps$, and $\TD=5)$, for the Ballet sequence model.} \label{table:comprison_exh}
\begin{tabular}{|c||c|c|c|c||c|c|c|c|} \hline
  &   \multicolumn{4}{|c|}{Static  Cameras}    &   \multicolumn{4}{|c|}{Moving Cameras}  \\   \cline{2-9}
  Optimization Method  &   \multicolumn{2}{c|}{$\rhos=0,\rhot=2$} & \multicolumn{2}{|c|}{$\rhos=2,\rhot=2$}  &   \multicolumn{2}{|c|}{$\rhos=0,\rhot=2$} & \multicolumn{2}{|c|}{$\rhos=2,\rhot=2$}  \\   \cline{2-9}
 & $K=3$ & $K=5$ & $K=3$ & $K=5$  & $K=3$ & $K=5$ & $K=3$ & $K=5$   \\ \hline \hline 
Exhaustive search algorithm &   24.39    &   24.54  & 26.50  & 26.65  &   23.13   &      23.19   &  25.07  &    25.20 \\ \hline
Branch pruning strategy &  24.39    &   24.52  &  26.47   &  26.63 &     23.11 &       23.16  &   25.05 &  25.18 \\ \hline
\end{tabular}
\end{center}
\end{table*}
%%%%%%%%%%%%%%%%%%%%%%%%
 %%%%%%%%%%%%%%%%%%%%%%%
%%% BREAKDANCER %%%%%%%%%%%
\begin{table*}[t]
\begin{center}
\caption{Average PSNR  of the reconstructed sequence for each camera  for systems with $4$ cameras   $(C=47\, Mbps$, $ r=23.5\, Mbps$, and $\TD=5)$, for the Breakdancer sequence model.} \label{table:Breakdancercomprison_exh}
\begin{tabular}{|c||c|c|c|c||c|c|c|c|} \hline
  &   \multicolumn{4}{|c|}{Static  Cameras}    &   \multicolumn{4}{|c|}{Moving Cameras}  \\   \cline{2-9}
  Optimization Method  &   \multicolumn{2}{c|}{$\rho_s=0,\rho_t=2$} & \multicolumn{2}{|c|}{$\rho_s=2,\rho_t=2$}  &   \multicolumn{2}{|c|}{$\rho_s=0,\rho_t=2$} & \multicolumn{2}{|c|}{$\rho_s=2,\rho_t=2$}  \\   \cline{2-9}
 & $K=3$ & $K=5$ & $K=3$ & $K=5$  & $K=3$ & $K=5$ & $K=3$ & $K=5$   \\ \hline \hline 
Exhaustive search algorithm  &   23.09     &    23.25    & 25.56     &  25.70   & 24.18      &  26.73         & 24.45    &   26.92    \\ \hline
Branch pruning strategy       &     23.03     &    23.23    &   25.54   &   25.67 &    24.18    &    26.70      &   24.43  &    26.91    \\ \hline
\end{tabular}
\end{center}
\end{table*}
%%%%%%%%%%%%%%%%%%%%%%%%%%%%%%%%%%%%%%%%%%%%%%

\begin{figure}[t]
\begin{center}
\subfigure[$C=23.5\, Mbps$]{
\includegraphics[width=3.1in,draft=false]{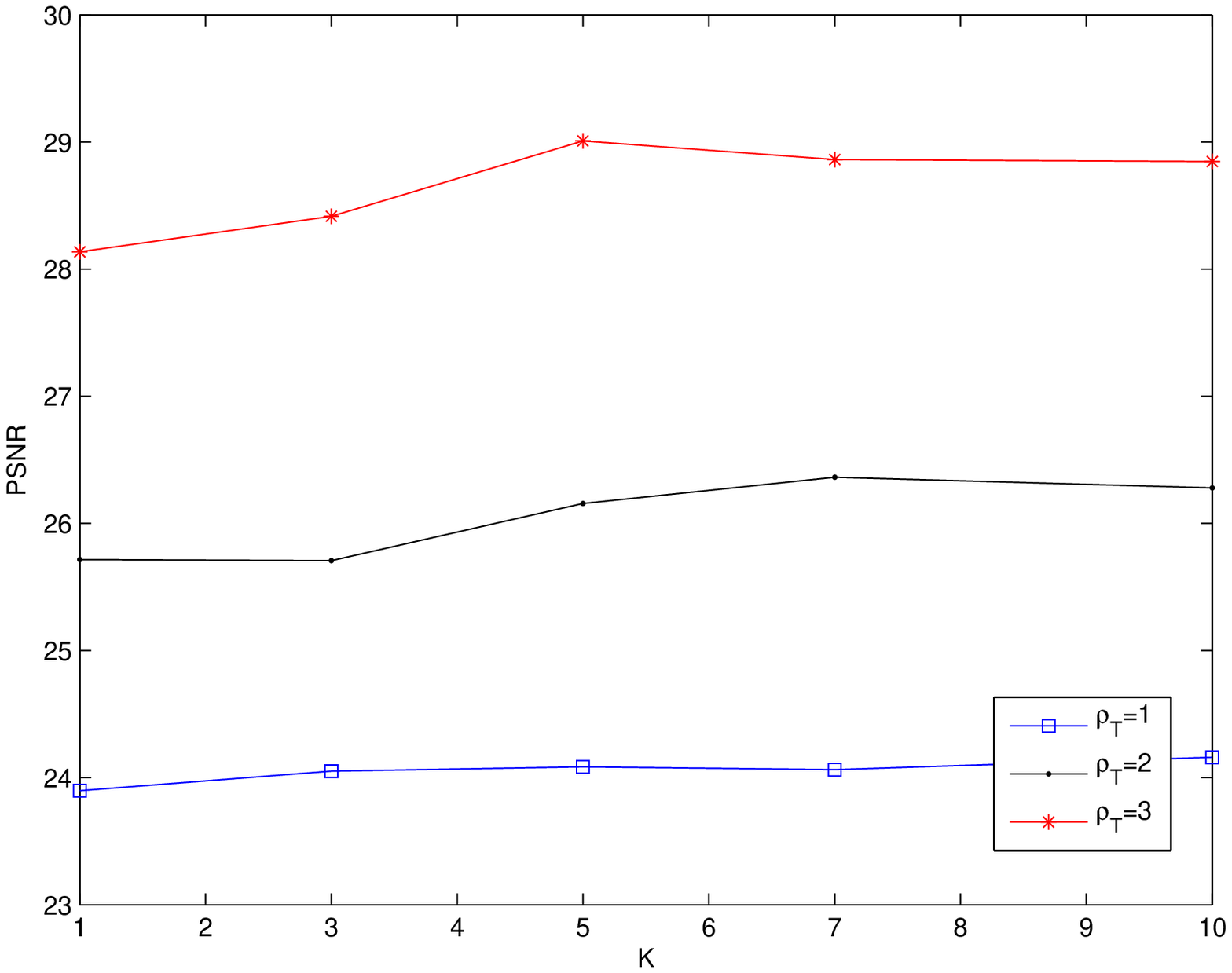}
\label{fig:experimental} }
 \subfigure[$C=47\, Mbps$]{
\includegraphics[width=3.1in,draft=false]{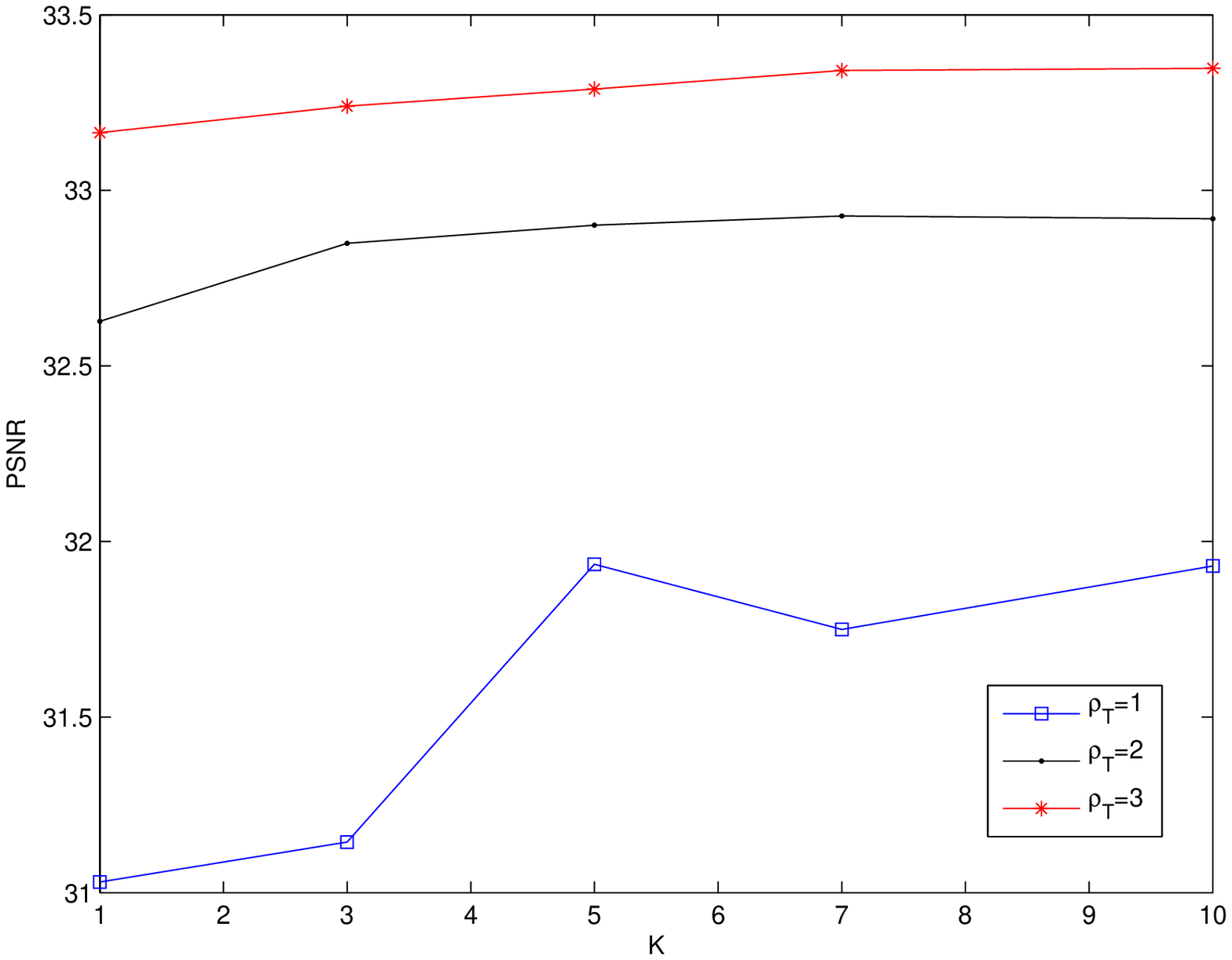}
\label{fig:theory}} 
\end{center}
\caption{PSNR vs optimization horizon $K$ for systems with $4$ dynamic
cameras ($r=23.5\, Mbps$, $\TD=5$,
  $\rho_{s}=2$, and $N_s=2$, Ballet model sequence).}
\label{fig:N4_R2_Corr2_PSNR.eps}
\end{figure}
%%%%%%%%%%%%%%%%%%%%%%%
%%% BREAKDANCER %%%%%%%%%%%
%%%%%%%%%%%%%%%%%%%%%%%
\begin{figure}[t]
\begin{center}
\subfigure[Ballet sequence model.]{
\includegraphics[width=3.1in,draft=false]{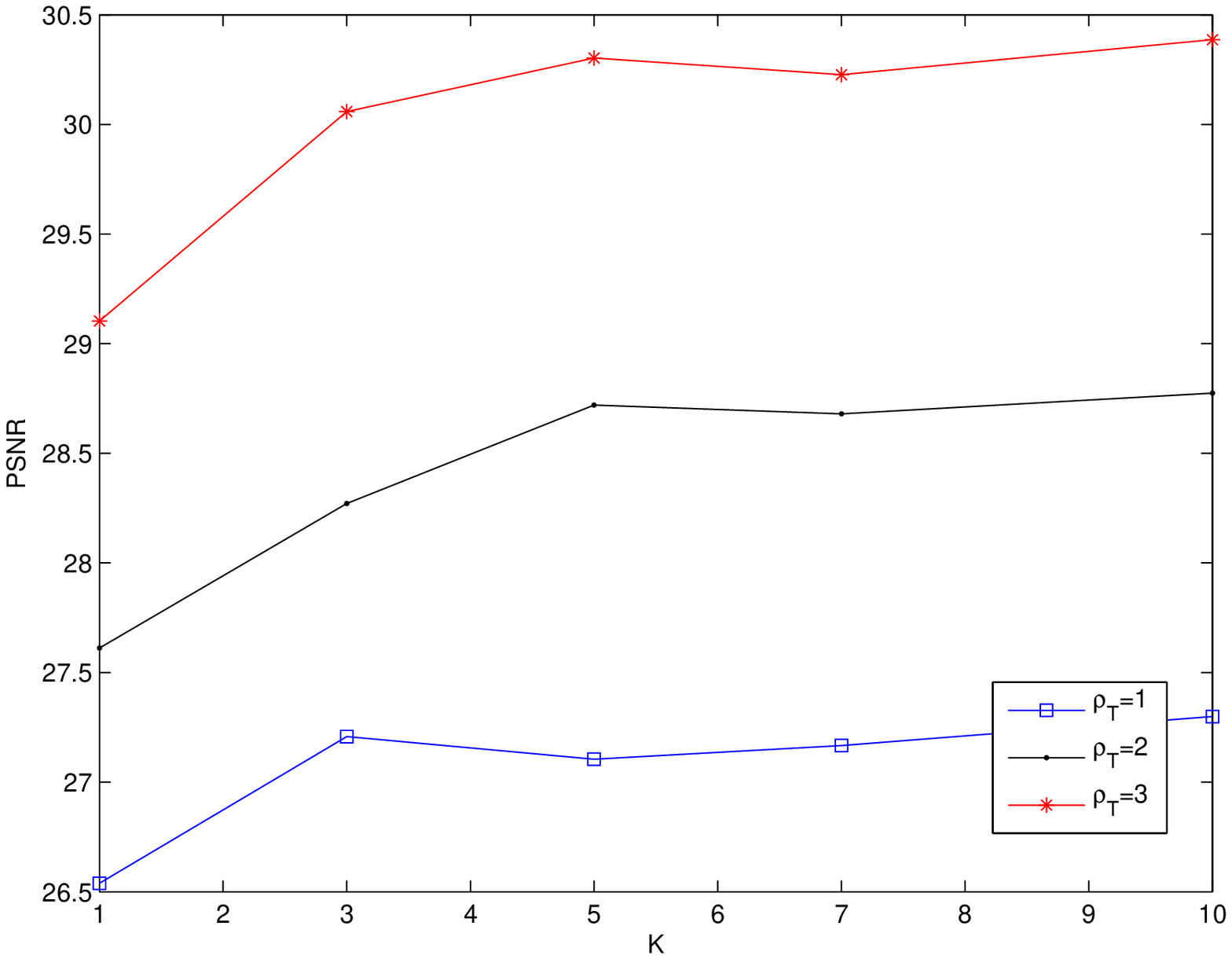}
\label{fig:DYNAMIC_N8_R200_Ta8_Corr4__1} }
 \subfigure[Breakdancer sequence model.]{
\includegraphics[width=3.1in,draft=false]{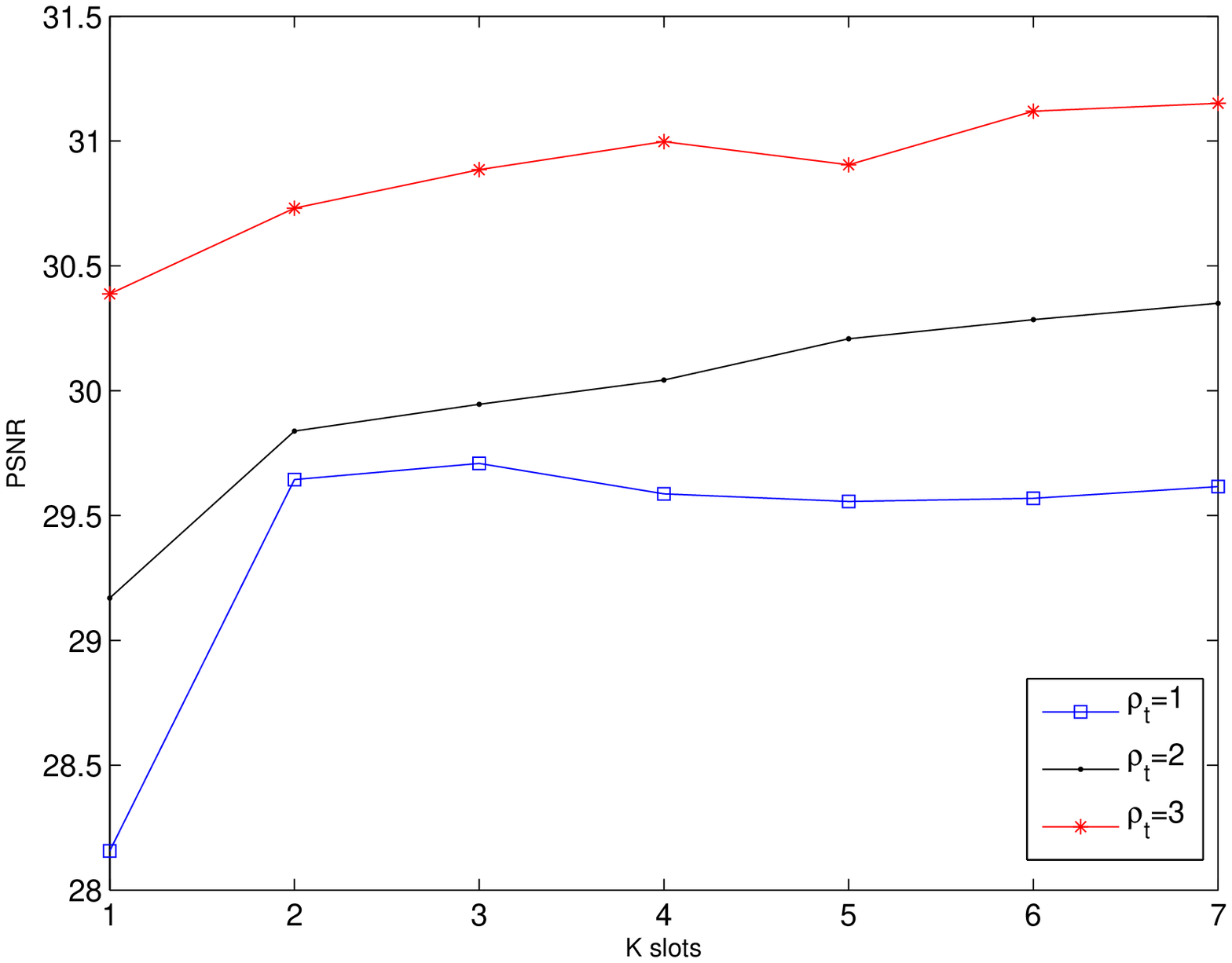}
\label{fig:Breakdancer_DYNAMIC_N8_R200_Ta8_Corr4}} 
\end{center}
\caption{Model-based reconstruction PSNR vs optimization horizon  $K$ for systems with $8$ dynamic
cameras ($C=47\, Mbps$, $r=23.5\, Mbps$, $\TD=5$,
 $\rho_{s}=4$,   and $N_s=2$).}
\label{fig:DYNAMIC_N8_R200_Ta8_Corr4}
\end{figure}

 \begin{figure}
  \begin{center}
\subfigure[Ballet sequence.]{
\includegraphics[width=3.1in,draft=false]{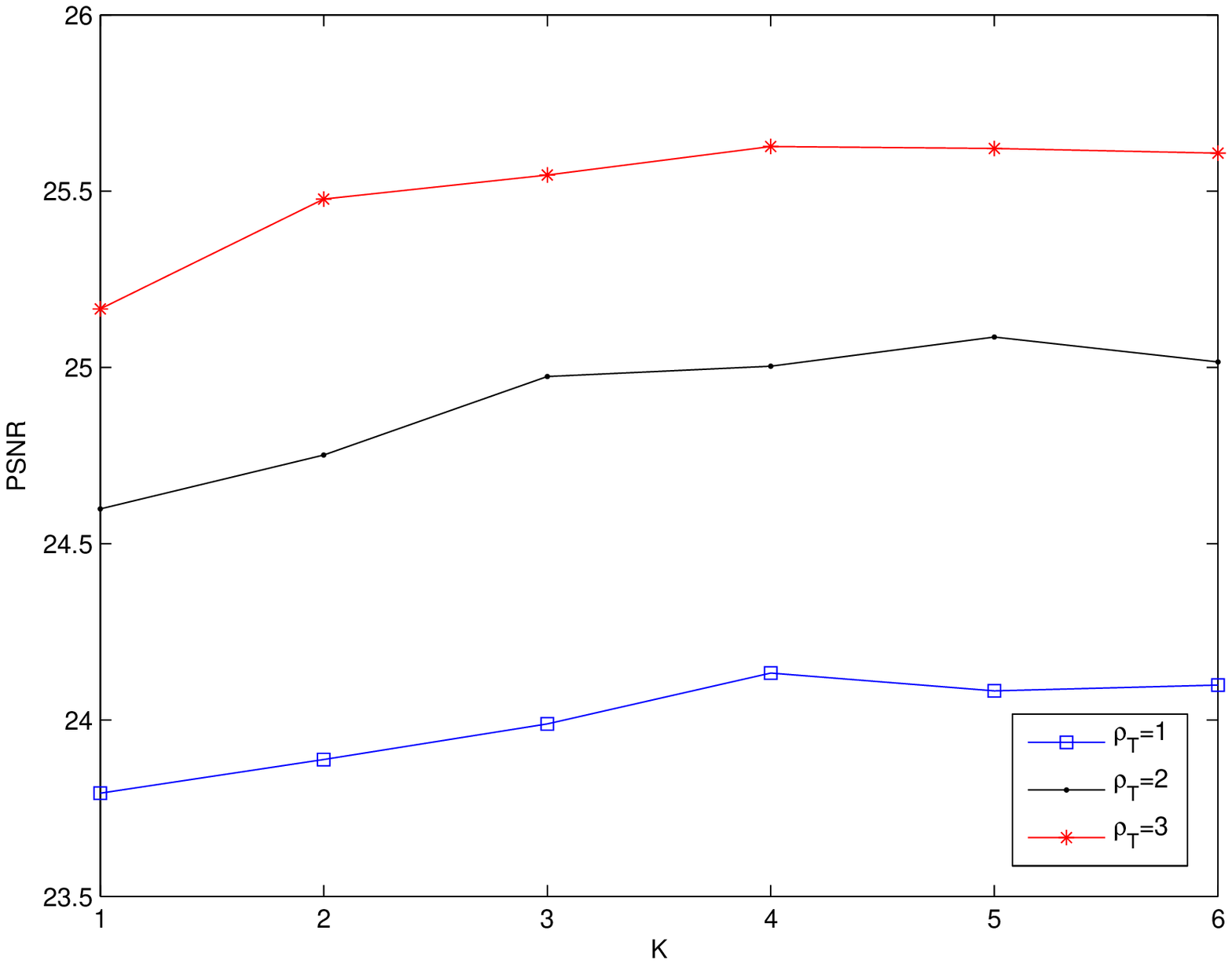}
\label{fig:Experimental_N8_MV1_Ta8__1} }
 \subfigure[Breakdancer sequence.]{
\includegraphics[width=3.1in,draft=false]{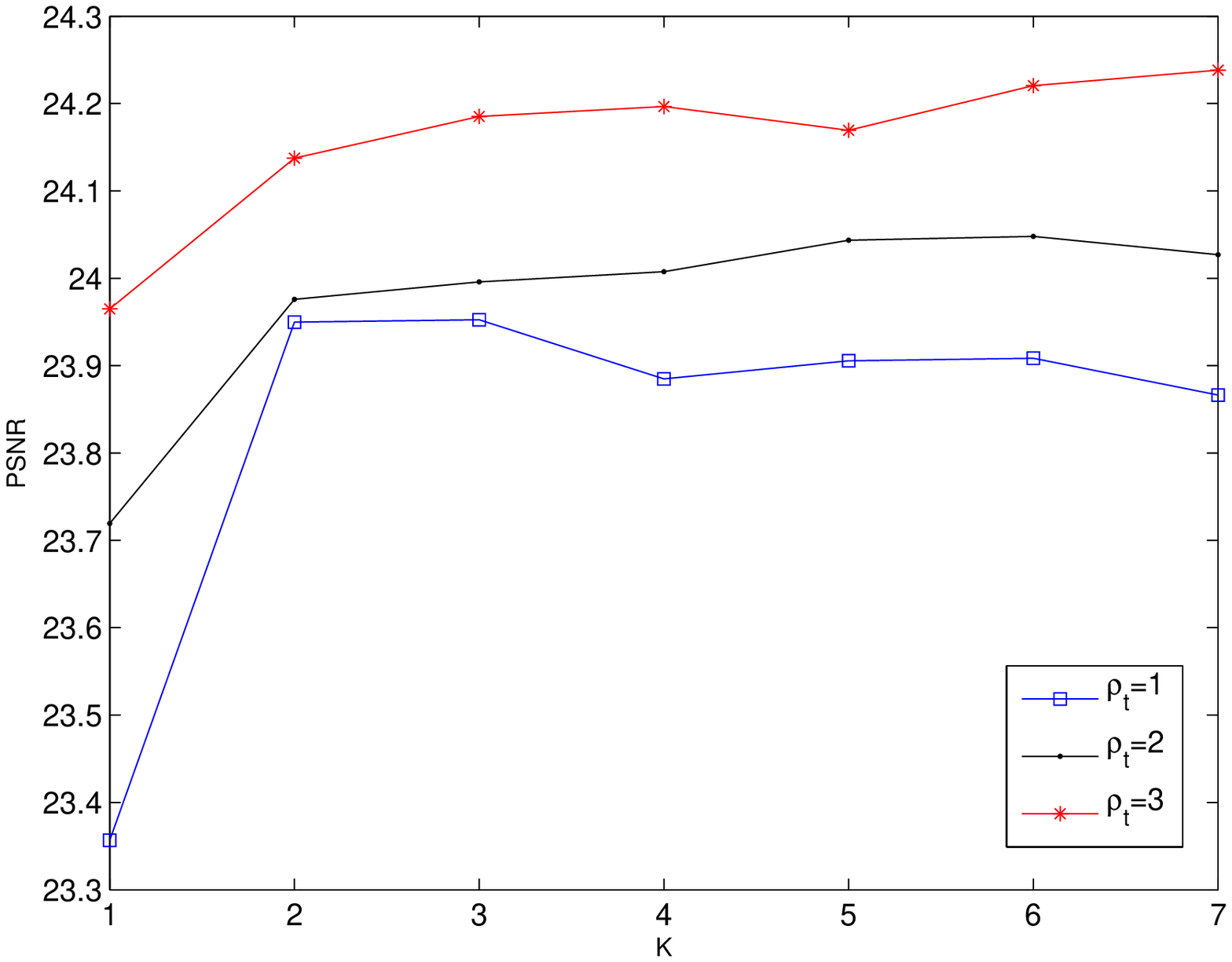}
\label{fig:Breakdancer_Experimental_N8_MV1_Ta8}} 
\end{center}
\caption{Reconstructed PSNR   for systems with $8$ dynamic
cameras ($C=47\, Mbps$, $r=23.5\, Mbps$, $\TD=5$,
  $\rho_{s}=4$, and $N_s=2$). }
\label{fig:Experimental_N8_MV1_Ta8}
\end{figure}

We first  compare the proposed sub-optimal scheduling algorithm with an optimal one. In particular, we randomly select a time instant $t\in[1,100]$  and  assume that  the scheduling history till the time instant $t-1$ is known \footnote{The scheduling history is randomly selected.}. We are interested in optimizing the scheduling policy over a time horizon of $K$ time slots with our trellis-based search technique and with an optimal solution, which exhaustively search for the best scheduling policy. Decoding  quality results for  the DUs acquired during the  time interval under consideration. Results of the reconstructed distortion of the DUs acquired during the  time instants $[1,t]$ are provided in Table \ref{table:comprison_exh} and in Table \ref{table:Breakdancercomprison_exh}, for the Ballet and Breakdancer video sequences, respectively. Each value is averaged over $1000$ random  simulations for both static and dynamic   scenarios  with $C=23.5\, Mbps$, $ r=11.7\, Mbps$, and $\TD=5$. It can be observed that, for both sequences,  the difference between   the branch pruning strategy and the exhaustive search method is negligible. This means that the pruning of the branches in the trellis-based optimization does  not penalize significantly the performance, while it drastically reduces the computational complexity.

%%%%%%%%%%%%%%%%%%%%%%%%
We now provide results for the proposed foresighted scheduling optimization in dynamic scenarios. In Fig.~\ref{fig:N4_R2_Corr2_PSNR.eps}, the model-based reconstructed PSNR is given   as a function of the number of optimization time slots $K$  for 
   systems with $4$ cameras for   several temporal correlation levels ($r=23.5\, Mbps$, $\rho_s=2$, $C=r $ and $C=2 r$).
 For all the temporal correlation values $\rho_t$, we  provide results for large $K$ and we  observe  performance gains  with $K$. Note that  the distortion gain due to large $K$ is sometimes marginal  for two main reasons:   i) the channel capacity is very limited and  only few DUs can be scheduled compared to the total number of acquired DUs (Fig.~\ref{fig:experimental} where the channel capacity is equal to the source   rate of one camera only);  
  ii) there are   large levels of correlation  so that the system performance is    less sensitive to non-optimal scheduling decisions   since most of the views will be reconstructed at a fair level anyway (see  Fig.~\ref{fig:theory} when $\rho_t=3$).

In Fig.~\ref{fig:DYNAMIC_N8_R200_Ta8_Corr4},  the PSNR quality is provided as a function of the optimization horizon $K$ for  systems with $8$ dynamic cameras ($C=47\, Mbps$, $r=23.5\, Mbps$, $\TD=5$,  and $\rho_{s}=4$) for both Ballet and Breakdancer video sequences. By increasing the number of cameras   from $4$ to $8$  but keeping   the ratio between the channel constraint $C$ and source rate $r$ constant, the number of DUs that cannot be scheduled increases; this makes    the selection of the best scheduling policy  even more crucial. As expected, the quality gain for large optimization horizons   gets more important in this case.

Finally, in Fig.~\ref{fig:Experimental_N8_MV1_Ta8},   experimental results are provided for systems with $8$ dynamic
cameras ($C=47\, Mbps$, $r=23.5\, Mbps$, $\TD=5$,
 and $\rho_{s}=4$). The experiment is the  same of  Fig.~\ref{fig:DYNAMIC_N8_R200_Ta8_Corr4} but the actual reconstruction of
 the scene is performed at the decoder. As already demonstrated for the greedy optimization results, the \emph{qualitative} behavior of the experimental and   model-based results is similar. In general we observe that, the larger the temporal correlation, the better the   quality  in the  reconstruction   since more past frames can be used in the reconstruction of a given frame. Furthermore, the experimental results confirm that increasing the optimization horizon improves the performance, as already observed in the results derived from the model-based results.

From the    simulation results, we can draw the following learnings.  First, we have demonstrated that the temporal and spatial correlations that  exist   among acquired frames in a multiview scenario is a crucial piece of information in the optimization of the streaming strategy. When packet filtering is imposed by bottleneck channels, the packet scheduling strategy  can drastically benefit from the  knowledge of the correlation  that exists between data units. We have also shown that a foresighted optimization strategy  outperforms greedy optimizations in most cases. Moreover, the benefit of considering the correlation level in the packet scheduling algorithm  increases in  dynamic   scenarios compared to static ones. The proposed algorithm is optimized in real time and refined at each transmission opportunity,  allowing to consider dynamic scenarios, in which  both   cameras positions and the level of correlation can vary in time. 
 In addition, it is worth noting that i)
 when the level of correlation exists   in both the time and space
 domains, knowing at least one of the two correlation levels leads to
   an improvement in the scheduling algorithm compared to the case
 where  no correlation information is known;   ii)  the knowledge of the
 correlation level   might help in selecting the best rate at which each
 camera should encode the images. In particular, the greater
 the level of correlation, the lower then number of views that needs to be  allocated per acquisition time for optimal performance.

Based on the above learnings, several possible research directions can be studied. The packet scheduling algorithm can be extended to  source coding optimization problems, where the rate of each view could be adapted over time. It could also be extended to scenarios with unreliable channels.  At large, the proposed framework  can be used in different systems in emerging   multiview video streaming applications, in which   both spatial and temporal correlations represent crucial information for adapting the video delivery solution.

%%%%%%%%%%%%%%%%%%%%%%%%%%%%%%%
\section{ {Conclusions}}
\label{sec:conclusions}
We have investigated the  impact of frame correlation for the scheduling
 of packets in a multi-camera system. In particular, we have proposed both a novel RD model able to take into account the correlation level among cameras and a method to estimate the contribution that each camera can offer in the reconstruction of correlated views. Based on this model,  
 we have proposed an optimization algorithm,   which determines  the packet scheduling policy by  taking into account the  channel capacity and   both the temporal and spatial correlations among    encoded frames.  The proposed   algorithm   is  able to adapt   the transmission strategy to the
level of correlation experienced by each camera. We have formalized a trellis-based optimization and we have proposed a suboptimal yet effective  solution with a tractable complexity, based on effective pruning in a trellis representation. 
Simulation results have demonstrated the gain of the proposed method compared to classical resource allocation techniques.
Finally, we have also demonstrated the robustness of foresighted optimization strategies.

\section*{Appendix \\ $\mat{\phi}$ Matrix Construction}
We now provide further details on the construction of the $\mat{\phi}$ matrix. The entire process is based on the subregions in which each frame is subdivided. In more details, the   number of regions  in which each frame is subdivided depends on both the video content and   the correlation level. Thus,   frames can be decomposed into different regions. Each region is designed by  a unique combination of correlated frames that are involved in the reconstruction at the decoder. 
In the temporal domain,   the contribution of neighboring frames to each region is evaluated by comparing images from the same camera. 
More precisely, each frame is subdivided into regions, each of them can be reconstructed from     previously  acquired frames only if no  motion occurs in these regions. As   no motion estimation  is employed at the source coding nor at the receiver in our system,   only the fixed background contributes to the   temporal extrapolation of missing frames.   
 In the spatial domain,  to evaluate the influence of each camera on the  neighboring ones,  
  we use DIBR techniques and calculate the number of pixels that can be estimated from neighboring views.

 \begin{figure} 
 \subfigure[ ]{
\includegraphics[width=3in,draft=false]{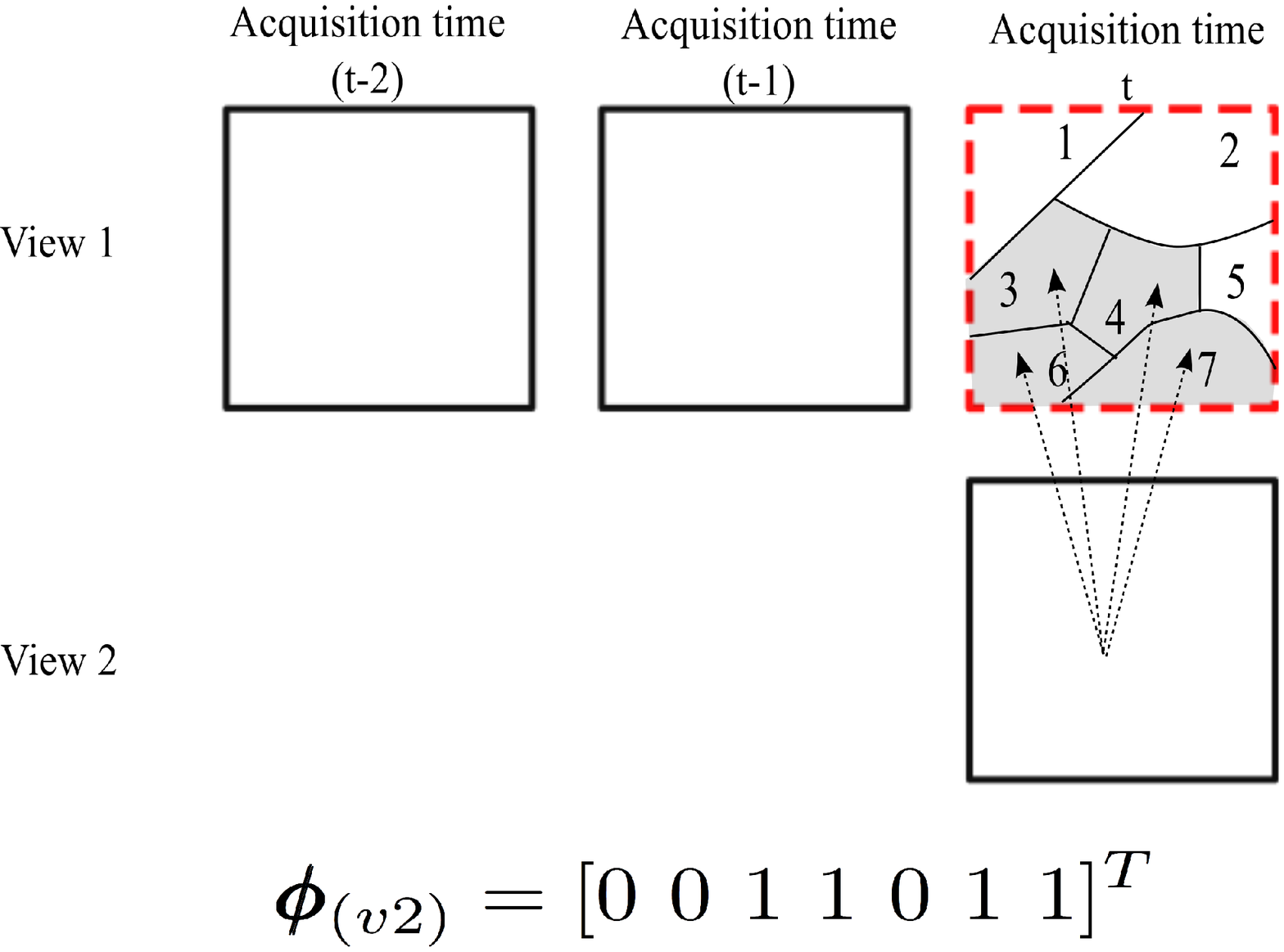}
\label{fig:From_v2}}
 \centering{
\subfigure[ ]{
\includegraphics[width=3in,draft=false]{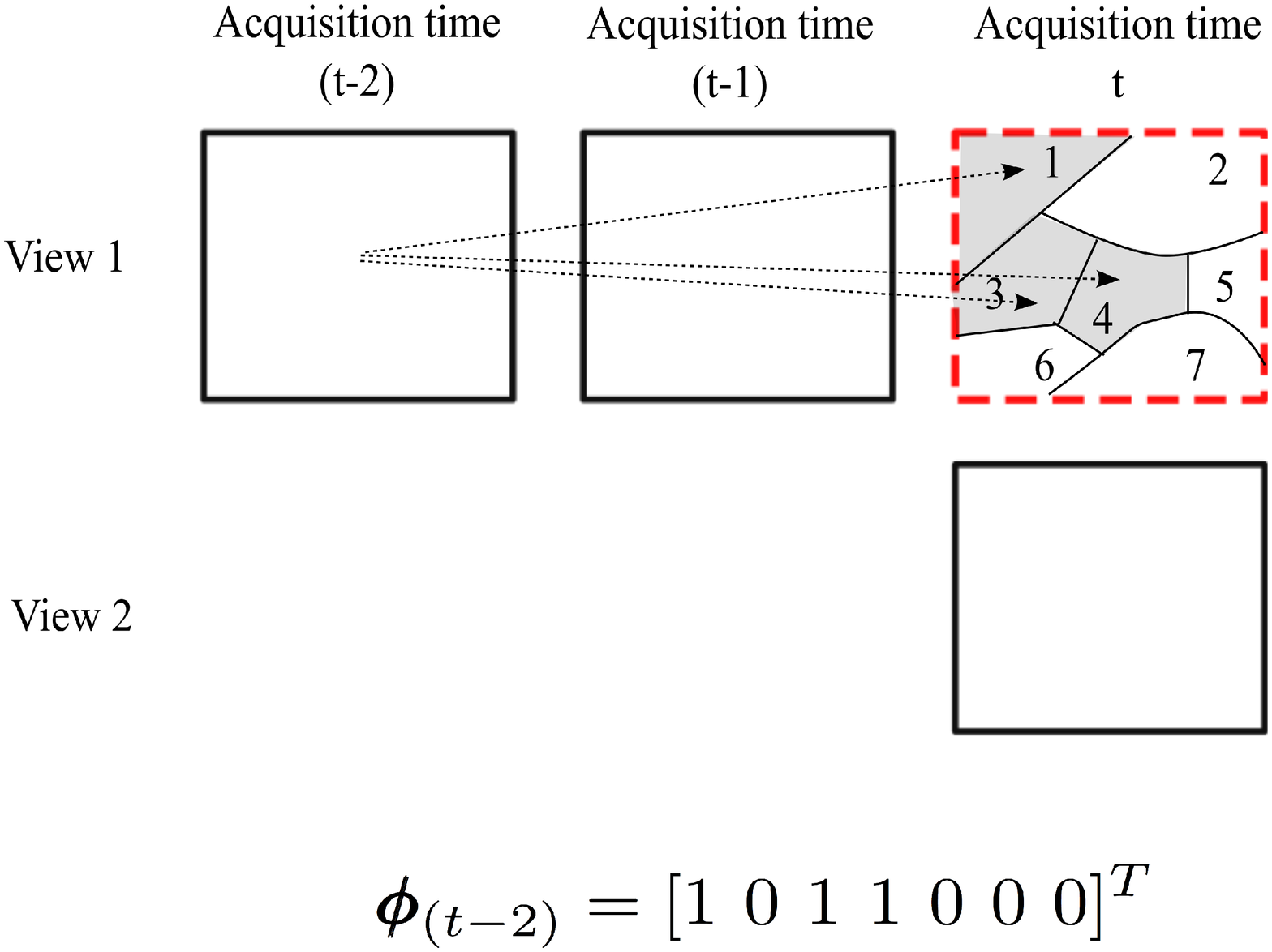}
\label{fig:From_t2} }}
 \centering{  \subfigure[ ]{
\includegraphics[width=3in,draft=false]{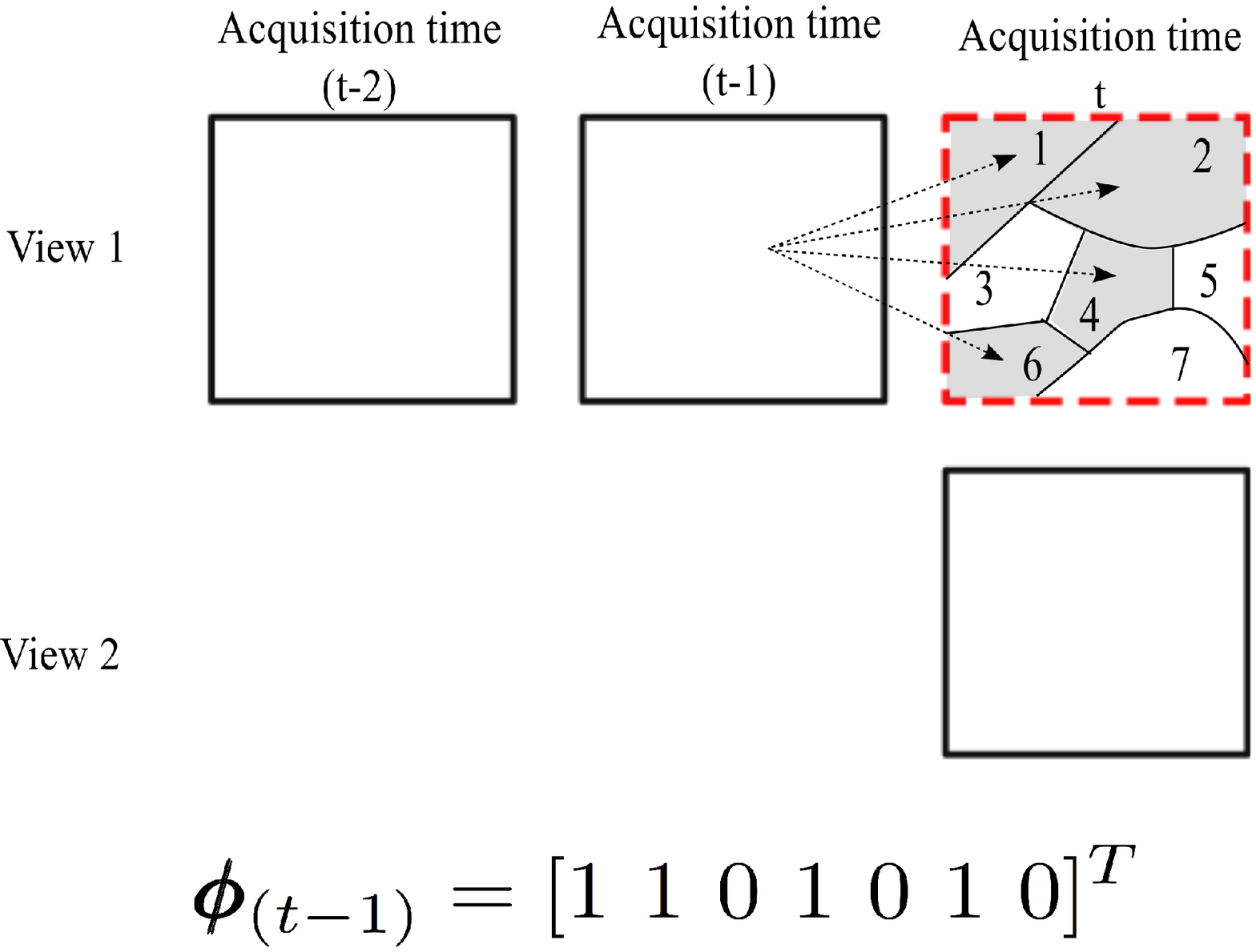}
\label{fig:From_t1}} }
\caption{Example in constructing the regions in which the frame $F_{t,1}$ is subdivided. (a) Regions of frame $F_{t,1}$ reconstructable from frame $F_{t,2}$.
(b) Regions of frame $F_{t,1}$ reconstructable from frame $F_{t-2,1}$. (c) Regions of frame $F_{t,1}$ reconstructable from frame $F_{t-1,1}$. }
\label{fig:example_phi}
\end{figure}

In Fig. \ref{fig:example_phi}, we provide an example to better explain how the $\mat{\phi}$ matrix is constructed and what is the meaning of each element of the matrix. In the figure, we consider a scenario in which two cameras acquire the   scene of interest. We are interested in evaluating the matrix for the  frame $F_{t,1}$ (depicted in the figure as a   dashed-border box), knowing that 
$F_{t,2}, F_{t-1,1}$ and $F_{t-2,1}$ are correlated to $F_{t,1}$. In Fig. \ref{fig:example_phi} (a)-(c) we show which parts of the frame 
$F_{t,1}$ can be reconstructed from $F_{t,2}, F_{t-1,1}$ and $F_{t-2,1}$, respectively. Regions that can be reconstructed (i.e., that are correlated) from the neighboring one are highlighted in grey in each subfigure. For example, from Fig. \ref{fig:example_phi} (a), we observe that regions   $3, 4, 6, $ and $7$ of frame  $F_{t,1}$ are correlated to  frame  $F_{t,2}$. Thus, we can build a vector $\mat{\phi}_{(v2)}=[0 \ 0Ê\ 1 \ 1\ 0 \ 1 \ 1]^T$ that maps the spatial correlation between the two views into the reconstructable subregions. 
If we then look at a specific region, say for example region $1$, we observe that the region is reconstructable from frame $F_{t-2,1}$  and $F_{t-1,1}$ but not from $F_{t,2}$. So, we can notice that each region has a unique combination of frames that can be used for the reconstruction at the decoder. Note also that all the regions of each frame can be always completely reconstructed from the same frame. This means that the vector $\mat{\phi}_{(v1)}$, which depicts the regions of $F_{t,1}$ that can be reconstructed from $F_{t,1}$,   corresponds to a unitary vector. Merging together all the vectors $\mat{\phi}_{(v1)}, \mat{\phi}_{(v2)}, \mat{\phi}_{(t-2)}, \mat{\phi}_{(t-1)}$, we obtain the following matrix
\begin{align}
\mat{\phi}_{1,t}& =[ \mat{\phi}_{(v1)} | \mat{\phi}_{(v2)} | \mat{\phi}_{(t-2)} | \mat{\phi}_{(t-1)}] = \begin{bmatrix} 
1 & 0 & 1 & 1 \\
1 & 0 & 0 & 1 \\
1 & 1 & 1 & 0 \\
1 & 1 & 1 & 1 \\
1 & 0 & 0 & 0 \\
1 & 1 & 0 & 1 \\
1 & 1  & 0 & 0
\end{bmatrix}\,.
\end{align}
Recalling the notation that we used in Sec. III,   $\mat{\phi}_{1,t}=\{\mat{\phi}_{j,1,t} \}_{j=1}^7$, where    $\mat{\phi}_{j,1,t}$ describes which of the neighboring frames can contribute to the reconstruction of the region $s_j$.
 From this matrix, we can notice that region $5$ can be reconstructed only if the frame    of interest (i.e., $F_{t,1}$) is received. Otherwise, the region cannot be reconstructed by any   neighboring frame. As general intuition, the larger is the level of correlation between frames, the greater is the number of non zero elements in the matrix.  
    
%%%%%%%%%%%%%%%%%%%%%%%%%%%%%%%
\bibliographystyle{IEEEtran}
\bibliography{DSC}
%%%%%%%%%%%%%%%%%%%%%%%%%%%%%%%%%

\end{document}